\documentclass[useAMS,usenatbib,twocolumn]{mnras}
\usepackage{graphics}
\usepackage{graphicx}
\usepackage{amsmath}
\usepackage{amssymb}
\usepackage{xspace}
\usepackage{commath}
\usepackage{times}
\usepackage{soul}
\usepackage{bm}
\usepackage{mathtools}
\usepackage{hyperref}
\hypersetup{colorlinks=true, linkcolor=black, citecolor=black, filecolor=blue, urlcolor=black}

\usepackage{acronym}
\newacro{PDF}{probability distribution function}
\newacroplural{PDF}[PDFs]{probability distribution functions}
\newcommand{\PDF}{\ac{PDF}}
\newcommand{\PDFs}{\acp{PDF}}
\newacro{DF}{distribution function}
\newacroplural{DF}[DFs]{distribution functions}
\newcommand{\DF}{\ac{DF}}
\newcommand{\DFs}{\acp{DF}}
\newacro{VRR}{vector resonant relaxation}
\newcommand{\VRR}{\ac{VRR}}
\newacro{SRR}{scalar resonant relaxation}
\newcommand{\SRR}{\ac{SRR}}
\newacro{NR}{non-resonant relaxation}
\newcommand{\NR}{\ac{NR}}
\newacro{BH}{black hole}
\newacroplural{BH}[BHs]{black holes}
\newcommand{\BH}{\ac{BH}}
\newcommand{\BHs}{\acp{BH}}
\newacro{IMBH}{intermediate mass black hole}
\newacroplural{IMBH}[IMBHs]{intermediate mass black holes}

\newcommand{\IMBHs}{\acp{IMBH}}

\newcommand{\rd}{\mathrm{d}}
\newcommand{\re}{\mathrm{e}}
\newcommand{\MBH}{M_{\bullet}}
\newcommand{\bK}{\mathbf{K}}
\newcommand{\bKp}{\mathbf{K}^{\prime}}
\newcommand{\Etot}{E_{\mathrm{tot}}}
\newcommand{\Ltot}{L_{\mathrm{tot}}}
\newcommand{\bLtot}{\mathbf{L}_{\mathrm{tot}}}
\newcommand{\Htot}{H_{\mathrm{tot}}}
\newcommand{\br}{\mathbf{r}}
\newcommand{\tp}{t^{\prime}}
\newcommand{\Min}{\mathrm{Min}}
\newcommand{\Max}{\mathrm{Max}}
\newcommand{\hbL}{\widehat{\mathbf{L}}}
\newcommand{\hL}{\widehat{L}}
\newcommand{\hLp}{\widehat{L}^{\prime}}
\newcommand{\hbLp}{\widehat{\mathbf{L}}^{\prime}}
\newcommand{\ellp}{\ell^{\prime}}
\renewcommand{\mp}{m^{\prime}}
\newcommand{\mJ}{\mathcal{J}}
\newcommand{\mH}{\mathcal{H}}
\newcommand{\half}{\tfrac{1}{2}}
\newcommand{\mO}{\mathcal{O}}
\newcommand{\p}{\partial}
\newcommand{\Fd}{F_{\mathrm{d}}}
\newcommand{\deltaD}{\delta_{\mathrm{D}}}
\newcommand{\Feq}{F_{\mathrm{eq}}}
\newcommand{\ellmax}{\ell_{\mathrm{max}}}
\newcommand{\kB}{k_{\mathrm{B}}}
\newcommand{\veps}{\varepsilon}
\newcommand{\bJ}{\mathbf{J}}
\newcommand{\Ndisc}{N_{\mathrm{disc}}}
\newcommand{\Npart}{N_{\mathrm{part}}}
\newcommand{\mmax}{m_{\mathrm{max}}}
\newcommand{\mmin}{m_{\mathrm{min}}}
\newcommand{\amax}{a_{\mathrm{max}}}
\newcommand{\amin}{a_{\mathrm{min}}}
\newcommand{\emin}{e_{\mathrm{min}}}
\newcommand{\emax}{e_{\mathrm{max}}}
\newcommand{\Myr}{\,\mathrm{Myr}}
\newcommand{\SR}{\mathrm{SR}}
\newcommand{\hLc}{\widehat{L}_{\mathrm{c}}}
\newcommand{\mc}{m_{\mathrm{c}}}
\newcommand{\thetac}{\theta_{\mathrm{c}}}
\newcommand{\fc}{f_{\mathrm{c}}}
\newcommand{\hbLdisc}{\widehat{\mathbf{L}}_{\mathrm{disc}}}
\newcommand{\bC}{\mathbf{C}}
\newcommand{\bT}{\bm{\theta}}
\newcommand{\ain}{a_{\mathrm{in}}}
\newcommand{\aout}{a_{\mathrm{out}}}
\newcommand{\ein}{e_{\mathrm{in}}}
\newcommand{\eout}{e_{\mathrm{out}}}
\newcommand{\Npop}{N_{\mathrm{pop}}}
\newcommand{\mD}{\mathcal{D}}
\newcommand{\hLz}{\widehat{L}_{z}}
\newcommand{\sinch}{\mathrm{sinch}}
\newcommand{\cotanh}{\mathrm{cotanh}}
\newcommand{\Mellk}{M_{\ell , k}}
\newcommand{\oF}{\overline{F}}
\newcommand{\bx}{\mathbf{x}}

\begin{document}

\title[Mass segregation and VRR]{
Orbital alignment and mass segregation in galactic nuclei\\
via vector resonant relaxation
}

\author[N. Magnan, J.-B. Fouvry, C. Pichon \& P.-H. Chavanis]{Nathan Magnan$^{1,2}$\thanks{nathan.magnan@maths.cam.ac.uk}, Jean-Baptiste Fouvry$^{1}$, Christophe Pichon$^{1,3}$
\newauthor
and Pierre-Henri Chavanis$^{4}$
\vspace*{6pt}\\
\noindent$^{1}$ CNRS and Sorbonne Universit\'e, UMR 7095, Institut d'Astrophysique de Paris, 98 bis Boulevard Arago, F-75014 Paris, France\\
\noindent$^{2}$ DAMTP, Centre for Mathematical Sciences, Wilberforce Rd, Cambridge CB3 0WA, United Kingdom\\
\noindent$^{3}$ Institut de Physique Th\'eorique, DRF-INP, UMR 3680, CEA, Orme des Merisiers B\^at 774, 91191 Gif-sur-Yvette, France\\
\noindent$^{4}$ Laboratoire de Physique Th\'eorique, Universit\'e de Toulouse, CNRS, UPS, France
}

\maketitle

\begin{abstract}
Supermassive black holes dominate the gravitational potential in galactic nuclei. In these dense environments, stars follow nearly Keplerian orbits and see their orbital planes relax through the potential fluctuations generated by the stellar cluster itself. For typical astrophysical galactic nuclei, the most likely outcome of this vector resonant relaxation (VRR) is that the orbital planes of the most massive stars spontaneously self-align within a narrow disc. We present a maximum entropy method to systematically determine this long-term distribution of orientations and use it for a wide range of stellar orbital parameters and initial conditions. The heaviest stellar objects are found to live within a thin equatorial disk. The thickness of this disk depends on the stars' initial mass function, and on the geometry of the initial cluster. This work highlights a possible (indirect) novel method to constrain the distribution of intermediate mass black holes in galactic nuclei. 
\end{abstract}
\begin{keywords}
Diffusion - Gravitation - Galaxies: kinematics and dynamics - Galaxies: nuclei
\end{keywords}

\section{Introduction}
\label{sec:intro}

Supermassive \BHs\ are ubiquitous in external galaxies~\citep{Magorrian+1998,Genzel+2010,Kormendy+2013} and their induced feedback plays a critical role in regulating galaxy formation through cosmic ages~\citep{Heckman+2014}. The unique proximity of our Galactic centre is an extraordinary opportunity to study and constrain the long-term evolution of galactic nuclei and the stellar clusters orbiting within. Recent developments in that realm include detailed census of stellar populations around SgrA*~\citep{Ghez+2008,Gillessen+2017}, the observation of a cool accretion disc~\citep{Murchikova+2019}, as well as the observation of the relativistic precession of S2~\citep{Gravity+2020}. Similarly, the origin of the observed clockwise stellar disc~\citep{Bartko+2009,Yelda+2014} has triggered a lot of interest, as its existence may impact the merger rate of the \IMBHs\ population~\citep{PortegiesZwart+2002}.

As already pointed in~\cite{Rauch+1996}, the steep potential well generated by the central \BH\ allows for efficient secular orbit-averaged interactions between stars, driving their relaxation through an intricate hierarchy of dynamical processes~\citep{Alexander2017}. In this paper, we focus on the process of \VRR\ during which stars undergo a random reshuffling of their orbital orientations through long-term coherent torques between the finite number of stellar orbital planes~\citep{Kocsis+2015}. Given that \VRR\ occurs on a timescale shorter than the stellar ages~\citep[see fig.~{1} in][]{Kocsis+2011}, one may expect that their observed distribution of orientations corresponds to some statistical equilibrium.

Determining the outcome of this long-term rearrangement has been the focus of recent efforts~\citep{Roupas+2017,Takacs+2018,Szolgyen+2018,Touma+2019,Tremaine2020II,Tremaine2020III,Gruzinov+2020} that jointly offer new clues on the fascinating properties of these long-range interacting systems, such as negative temperatures or phase transitions. In particular,~\cite{Szolgyen+2018}, using an ingenious Monte--Carlo approach, have shown that in systems with a wide range of stellar populations (i.e.\ various masses and semi-major axes), \VRR\ can lead to the spontaneous formation of a disc through the angular segregation of the most massive stars and \IMBHs\@.

This is the issue that we further investigate in this work.
We develop and implement an explicit and efficient optimisation procedure to find maximum entropy solutions compatible with an initial configuration, so as to infer the thermodynamical equilibria of a given galactic nucleus. In the particular context of \VRR\@, this roadmap was already started out by~\cite{Roupas+2017} in the limit of a single-population system with a quadrupolar interaction, and later improved in~\cite{Takacs+2018} which, while still restricting themselves to a single-population system, considered harmonic expansion beyond the quadrupole. Here, we expand these works to multi-population clusters by emphasising the critical role played by the diversity of stellar orbits and masses to allow for non-trivial anisotropic statistical equilibria, as unveiled in~\cite{Szolgyen+2018}.

The paper is organised as follows. In~\S\ref{sec:model}, we briefly review the process of \VRR\ in galactic nuclei. We also present the numerical optimisation method used to determine the thermodynamical equilibria. In~\S\ref{sec:application}, we use this approach to carefully investigate the details of equilibrium configurations that appear in systems composed of multiple stellar populations, as well as the impact of the initial mass function and the geometry of the formation scenario. Finally, we conclude in~\S\ref{sec:conclusion}.

\section{VRR model}
\label{sec:model}

Let us first set up the framework in which to quantify the long-term effect of \VRR\ on a population of orbital planes.

\subsection{Interaction of Keplerian annuli via VRR}

We consider an isolated cluster of ${ N \!\gg\! 1 }$ stars orbiting a supermassive \BH\ of mass $\MBH$. Following a double orbit-average over all the stars' Keplerian motion and in-plane precession, the \VRR\ Hamiltonian takes the form\footnote{We also assume that ${ \sum_{i=1}^{N}m_{i} \!\ll\! \MBH }$~\citep{Roupas2020}.}~\citep{Kocsis+2015}
\begin{equation}
	\Htot = - \sum_{i < j}^{N} \bigg\langle \frac{G m_{i} m_{j}}{|\br_{i} (t) - \br_{j} (\tp)|} \bigg\rangle_{t , \tp} ,
	\label{def:Htot}
\end{equation}
where the sum over ${ (i,j) }$ runs over all the pairs of particles, and the double orbit-average ${ \langle \, \cdot \, \rangle_{t , \tp} }$ operates over the fast Keplerian motions and in-plane precessions of both particles, with ${ \br_{i} (t) }$ and ${ \br_{j} (\tp) }$ describing their trajectories.

\begin{figure}
	\centering
	\includegraphics[width=0.45 \textwidth]{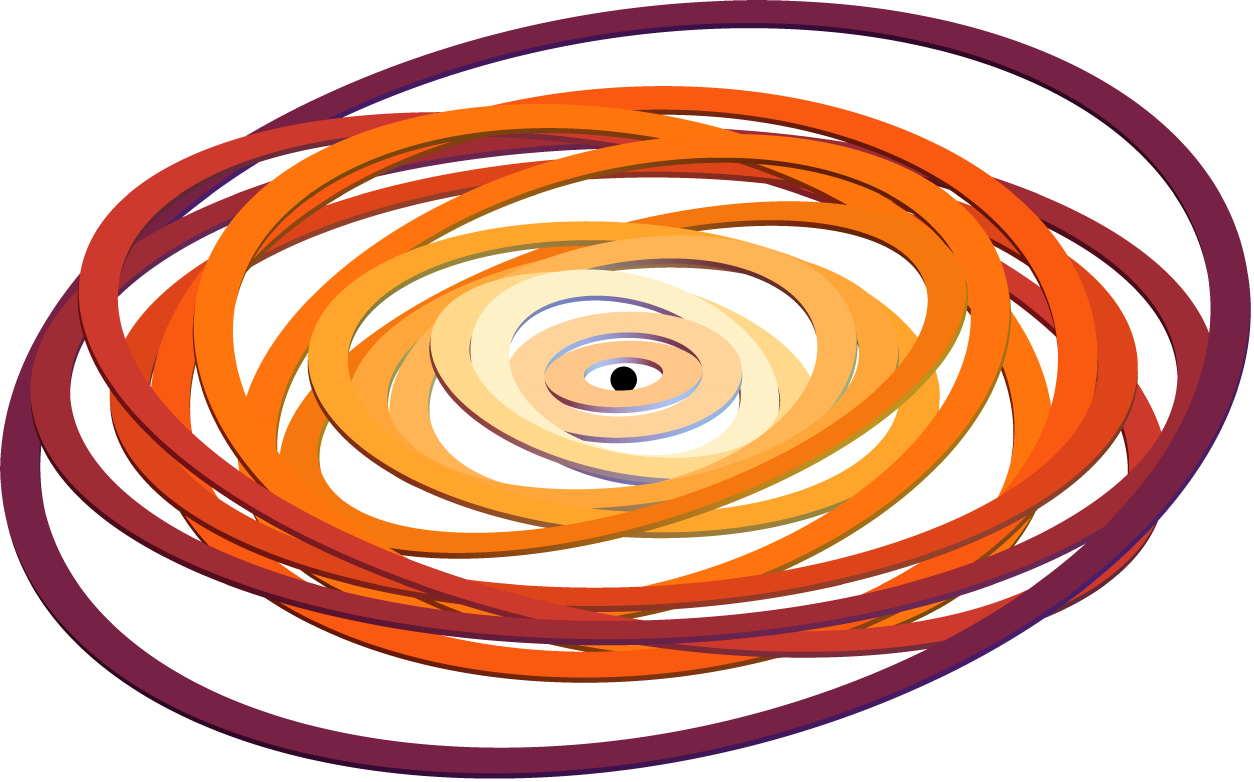}
	\caption{Illustration of the process of \VRR\ in a galactic nucleus. Following an orbit-average over the fast Keplerian motion induced by the central \BH\ and the in-plane precessions (due to both the stellar mean potential and the relativistic corrections), stars are replaced by massive annuli (here quasi-circular) which can torque one another \citep[see also fig.~{1} in][]{Giral+2020}. This leads ultimately to the relaxation of the stellar orbital orientations with possibly the spontaneous formation of aligned discs.}
	\label{fig:Annuli}
\end{figure}

Using the Legendre expansion of the Newtonian interaction, and the addition theorem for spherical harmonics, one can rewrite equation~\eqref{def:Htot} as~\citep[see, e.g.\@,][]{Fouvry+2019}
\begin{align}
	\Htot {} &= - \sum_{i < j}^{N} \sum_{\ell = 2}^{\ellmax} \sum_{m = - \ell}^{\ell} \!\! \mH_{\ell} \big[ \bK_{i} , \bK_{j} \big] \, Y_{\ell m} (\hbL_{i}) \, Y_{\ell m} (\hbL_{j}) ,
	\label{eq:Htot}
\end{align} 
where ${ \hbL_{i} }$ stands for the unit vector aligned with the instantaneous orientation of the star's orbital plane and the real spherical harmonics, ${ Y_{\ell m} (\hbL) }$, are normalised so that ${ \!\int\! \rd \hbL \, Y_{\ell m} Y_{\ellp \mp} \!=\! \delta_{\ell\ellp}\delta_{m \mp}}$.
In equation~\eqref{eq:Htot}, a Keplerian annulus (see Fig.~\ref{fig:Annuli}) is fully characterised by its conserved quantities
\begin{equation}
	\bK = (m , a , e) ,
	\label{def:K}
\end{equation}
with $m$ the star's individual mass, $a$ its semi-major axis, and $e$ its eccentricity.
We stress that, owing to orbit average, each $\bK_{i}$ are taken to be fixed throughout the \VRR\ dynamics. Finally, in equation~\eqref{eq:Htot}, we introduced the (symmetric) coupling coefficient
\begin{align}
	\mH_{\ell} \big[ \bK_{i} , \bK_{j} \big] = 
		{} & G m_{i} m_{j} \, \frac{4 \pi }{2 \ell \!+\! 1} \big| P_{\ell} (0) \big|^{2} \nonumber
	\\
		{} & \times \int_{0}^{\pi} \!\! \frac{\rd M_{i}}{\pi} \!\! \int_{0}^{\pi} \!\! \frac{\rd M_{j}}{\pi} \, \frac{\Min [r_{i} , r_{j}]^{\ell}}{\Max [r_{i} , r_{j}]^{\ell + 1}} ,
	\label{def:J_l}
\end{align}
with $P_{\ell}$ the Legendre polynomial of order $\ell$ and $M_{i}$ the mean anomaly of the orbit $\bK_{i}$~\citep[see, e.g.\@,][]{MurrayDermott1999}. These coupling coefficients fully encode the impact of the shape of the two orbits on their relative torque and show significant diversity~\citep[see, e.g.\@, fig.~{1} in][]{Kocsis+2015}\footnote{Following~\S{A} of~\cite{Fouvry+2019}, the present coupling coefficients simply read ${ \mH_{\ell} [\bK , \bKp] \!=\! \mJ_{\ell} [\bK , \bKp] \, L (\bK) }$, with ${ L (\bK) }$ introduced in equation~\eqref{def_LK}. In practice, following the notations from~\S{A} of~\cite{Fouvry+2019}, the coupling coefficients were pre-computed once, up to ${ \ellmax = 50 }$ and on a ${ 200 \!\times\! 100 \!\times\! 100 }$ linear grid in ${ (\ln(\ain/\aout),\ein,\eout) }$ with ${ 10^{-2} \!\leq\! \ain/\aout \!\leq\! 1 }$ and ${ 0 \!\leq\! \ein , \eout \!\leq\! 0.99 }$, to be subsequently interpolated.}.

Finally, we point out a few specificities of the multipole expansion from Eq.~\eqref{eq:Htot}: (i) All odd $\ell$ harmonics have vanishing coupling coefficients, ${ \mH_{\ell} \!=\! 0 }$, so that they can be dropped; (ii) The Legendre expansion is truncated to the maximum order ${ \ell \!\leq\! \ellmax }$; (iii) We do not account for the harmonics ${ \ell \!=\! 0 }$ as it does not drive any dynamics. With such a convention, spherically symmetric distributions have a vanishing mean total energy.

Describing the process of \VRR\ amounts to describing the long-term evolution of $\hbL_{i}$, as driven by
the Hamiltonian from equation~\eqref{eq:Htot}~\citep[see~\S{2.2} in][for the associated equations of motion]{Kocsis+2015}. We illustrate such a system in Fig.~\ref{fig:Annuli}. One can equivalently describe the instantaneous state of the stellar cluster with its discrete \DF\
\begin{equation}
	\Fd (\hbL , \bK , t) = \sum_{i = 1}^{N} \deltaD (\hbL - \hbL_{i} (t)) \, \deltaD (\bK - \bK_{i}) ,
	\label{def:Fd}
\end{equation}
which follows the normalisation ${ \!\int\! \rd \hbL \rd \bK \Fd \!=\! N }$. Assuming that particles with the same orbital parameters are indistinguishable, this \DF\ entirely describes the system's state. Therefore, characterising the \VRR\ dynamics amounts to characterising ${ \p \Fd / \p t }$, using the Klimontovitch equation~\citep{Klimontovich1967}.

\subsection{Equilibrium configurations}

Fortunately, the \VRR\ dynamics (${ \sim\! 1\Myr }$ for S2 around SgrA*, see fig.~{1} in~\cite{Kocsis+2011}) is rapid compared to the cluster's age (${ \sim\! 10\Myr }$ for the S-cluster, see~\cite{Habibi+2017}). As such, if one is interested in sufficiently long timescales, rather than describing the details of ${ \p \Fd / \p t }$, one may solely focus on characterising the expected equilibrium configurations reached at late times. This is given by
\begin{equation}
	\Feq (\hbL , \bK) = \lim\limits_{t \to + \infty} \big\langle \Fd (\hbL , \bK , t) \big\rangle ,
	\label{def_Feq}
\end{equation}
with ${ \langle \, \cdot \, \rangle }$ standing for an ensemble average over independent realisations of the system -- see~\S\ref{sub:AxiAssump} for the appropriate handling of this ensemble average. Efficiently predicting these long-term equilibrium \DFs\ is the focus of the present work.

In such a late-time limit, the only information retained by the cluster are its invariants.
These are:
\begin{itemize}
	\item The number density of stars with orbital parameters $\bK$
	\begin{equation}
		N (\bK) = \!\! \int \!\! \rd \hbL \, \Feq (\hbL , \bK) .
		\label{def_NK}
	\end{equation}
	\item The total energy
	\begin{equation}
		\Etot = \half \!\! \int \!\! \rd \hbL \rd \bK \, \Feq (\hbL , \bK) \, \veps (\hbL , \bK) ,
		\label{def_Etot}
	\end{equation}
	where ${ \veps (\hbL , \bK) }$ stands for the energy of a particle of orientation $\hbL$ and orbital parameter $\bK$ as
	\begin{equation}
		\veps (\hbL , \bK) \!=\! - \!\! \sum_{\ell = 2}^{\ellmax} \! \sum_{m = - \ell}^{\ell} \! \int \!\! \rd \bKp \, \mH_{\ell} \big[ \bK , \bKp \big] M_{\ell m} (\bKp) Y_{\ell m} (\hbL) ,
		\label{def_veps}
	\end{equation}
	with the magnetisations
	\begin{equation}
		M_{\ell m} (\bK) = \!\! \int \!\! \rd \hbL \, Y_{\ell m} (\hbL) \, \Feq (\hbL , \bK)
		\label{def_M}
	\end{equation}
	defined as spherical harmonic moments of the \DF\@. 
	\item The total angular momentum
	\begin{equation}
		\bLtot = \!\! \int \!\! \rd \hbL \, \rd \bK \, L (\bK) \, \hbL \, \Feq (\hbL , \bK) ,
		\label{def_Ltot}
	\end{equation}
	with the norm of the angular momentum vector
	\begin{equation}
		L (\bK) = m \sqrt{G \MBH \, a \, (1 - e^{2})} .
		\label{def_LK}
	\end{equation}
\end{itemize}

As usually carried out in the microcanonical ensemble (i.e.\ for an isolated cluster), for a given set of invariants ${ \{ N (\bK) , \Etot , \bLtot \} }$, the admissible equilibrium configurations are obtained by maximising the Boltzmann entropy
\begin{equation}
	S = - \kB \!\! \int \!\! \rd \hbL \rd \bK \, \Feq (\hbL , \bK) \, \ln \!\big[ \Feq (\hbL , \bK) \big] ,
	\label{def_S}
\end{equation}
with $\kB$ the Boltzmann constant,
under the previous conservation constraints. Such a maximisation generically yields (see~\S\ref{sec:Feq}) 
\begin{align}
	\Feq(\hbL, \bK) = N(\bK) \, \frac{ \re^{- \beta \, \veps (\hbL , \bK) + L (\bK) \, \bgamma \cdot \hbL}}{
	\displaystyle \int \!\! \rd \hbLp \, \re^{- \beta \, \veps (\hbLp , \bK) + L (\bK) \, \bgamma \cdot \hbLp}} ,
	\label{eq:Shape_from_Lagrange}
\end{align}
where $\beta$ and $\bgamma$ are the Lagrange multipliers respectively associated with the total energy and angular momentum conservation. We note that equation~\eqref{eq:Shape_from_Lagrange} is very similar to its co-planar counterpart found in the HMF model~\citep{Antoni+1994,Chavanis+2005} to capture bar formation~\citep{Pichon+1993} as a phase transition towards orbit alignment.

\cite{Gruzinov+2020} recently used a similar
mean-field approximation and maximum entropy method
to determine the thermodynamical equilibria
of black hole star clusters.
It is therefore no surprise that their equation~{(3)}
is so similar to the present equation~\eqref{eq:Shape_from_Lagrange}.
The main difference is that~\cite{Gruzinov+2020}
considered the case of massive Keplerian elliptic wires,
while we consider the case of massive Keplerian annuli.
Phrased differently, \cite{Gruzinov+2020} investigated
the simultaneous equilibria of \SRR\ and \VRR\
-- i.e.\ the joint relaxation of eccentricities and orientations --
whereas we focus here on the equilibria of \VRR\
-- i.e.\ the sole relaxation of orientations.

\subsection{Axisymmetric assumption}
\label{sub:AxiAssump}

For a given nuclear cluster, the total angular momentum vector $\bLtot$ provides us with only one specific direction, taken to be the ${ + z }$ axis throughout the paper. Unfortunately, this does not imply that the thermodynamical equilibrium of \VRR\ necessarily has an axial symmetry around $\bLtot$, as a spontaneous symmetry breaking could occur~\citep[see, e.g.\@,][]{Kocsis+2011,Gruzinov+2020}.

For simplicity however, we assume that \VRR\ does not exhibit here any such symmetry breaking, and restrict ourselves to axisymmetric \DFs\@, i.e.\ 
\begin{equation}
	\forall \, \bK, \, \forall \, \ell, \, \forall \, m \neq 0, \, M_{\ell m} (\bK) = 0 .
	\label{def:axisymmetry}
\end{equation}
From there, we can make the simplifications ${ \bLtot \!\to\! \Ltot \!>\! 0 }$ and ${ \bgamma \!\to\! \gamma \!>\! 0 }$. This assumption greatly reduces the total number of spherical harmonics to consider, therefore it significantly alleviates the numerical complexity. However, let us stress that this assumption is not always physically motivated. It is still legitimate in some cases, as for single-population and single-harmonic clusters~\citep{Roupas+2017} or multi-population clusters with null inverse temperatures (\S\ref{sec:beta_zero}), but it does not hold in some other regimes, as emphasised by the finding of a warped \VRR\ disc in the simulations of~\cite{Kocsis+2011} (see fig.\@~{6} therein). Overall, the axisymmetric assumption is an important limitation of the present work, and needs to be challenged in future studies.

To comply with this approach, the ensemble average of equation~\eqref{def_Feq} is carried out over realisations which are all rotated to have their $\bLtot$ aligned along ${ +z }$.

\subsection{Self-consistency}
\label{sub:SelfCons}

Of course, one needs to impose self-consistency on equations~\eqref{def_Etot} and~\eqref{def_Ltot}, as well as on equation~\eqref{eq:Shape_from_Lagrange}. Indeed, ${ \Feq (\hbL , \bK) }$ involves the one-particle energy ${ \veps (\hbL , \bK) }$ which, via equations~\eqref{def_veps} and~\eqref{def_M}, involves ${ \Feq (\hbL , \bK) }$ itself. Within the present microcanonical ensemble, imposing the cluster's total energy, angular momentum and orbital distribution ultimately sets up its temperature (via $\beta$), rate of rotation (via $\gamma$), and shape (via ${ M_{\ell 0} (\bK) }$)\footnote{In principle, there are infinitely many harmonics $\ell$, but in practice we found it acceptable to stop at ${ \ellmax \!=\! 10 }$, see~\S\ref{sec:Convlmax}.}.

To effectively solve such a generic problem, we discretise the distribution as a finite set of stellar populations, indexed by $k$ and described by the orbital parameters $\bK_{k}$. We then write ${ \Feq (\hbL , \bK) \!=\! \sum_{k} \! F_{k} (\hbL) \, \deltaD (\bK \!-\! \bK_{k}) }$ with ${ F_{k} (\hbL) }$ the distribution of orientation of the $k^{\mathrm{th}}$ population and ${ N_{k} \!=\! \!\int\! \rd \hbL F_{k} (\hbL) }$ its number of stars (see equation~\ref{def_NK}). After this discretisation, a cluster's configuration is fully characterised by its set of order parameters
\begin{equation}
	\bT = \big( \beta , \gamma , \{ \Mellk \} \big) ,
	\label{def_bT}
\end{equation}
with ${ \Mellk \!=\! \!\int\! \rd \hbL Y_{\ell 0} (\hbL) F_{k} (\hbL) }$
the axisymmetric $\ell$-magnetisation of the $k^{\mathrm{th}}$ population (see equation~\ref{def_M}).

For a given initial condition, i.e.\ a given ${ (\Etot , \Ltot , \{ N_{k} \} ) }$, the cluster's equilibrium configurations are obtained for the parameters $\bT$ that are joint roots of the consistency functions
\begin{align}
C_{E} {} & = \big( \Etot - \Etot \big[ \Feq (\bT) \big] \big) / \Etot ,
\nonumber
\\
C_{L} {} & = \big( \Ltot - \Ltot \big[ \Feq (\bT) \big] \big) / \Ltot ,
\nonumber
\\
\forall \, \ell , \, \forall \, k , \, C_{\Mellk} {} & = \big( \Mellk \!-\! \Mellk \big[ \Feq (\bT) \big] \big) / ( N_{k} \, y_{\ell} ) ,
\label{def_consistency}
\end{align}
with ${ y_{\ell} \!=\! \sqrt{(2 \ell \!+\! 1)/(4 \pi)} }$
and ${ \Feq (\bT) }$ coming from equation~\eqref{eq:Shape_from_Lagrange}. Importantly, in order to place all the constraints from equation~\eqref{def_consistency} on equal footings, the consistency functions ${ \bC \!=\! \big( C_{E} , C_{L} , C_{\Mellk} \big) }$ are all dimensionless, and rescaled to be of order unity.

In order to find roots of the function ${ \bT \!\mapsto\! \bC [\bT] }$, we use Newton--Raphson's method~\citep[e.g.\@,][]{Press+2007}, following~\cite{Takacs+2018}. More precisely, starting from a configuration $\bT_{n}$, we compute
\begin{equation}
	\bT_{n+1} = \bT_{n} - \bJ^{-1} [\bT_{n}] \, \bC [\bT_{n}] ,
	\label{iteration_Newton}
\end{equation}
to obtain the next iteration\footnote{Note that in practice, it is faster and numerically more stable to solve the linear equation ${ \bJ \, \bx = - \bC }$ for the unknown ${ \bx \!=\! \bT_{n+1} \!-\! \bT_{n} }$, rather than to explicitly compute the inverse matrix, $\bJ^{-1}$.}, with ${ \bJ [\bT]_{i j} \!=\! \p C_{i} / \p \theta_{j} }$  the Jacobian of the consistency function $\bC$ at point $\bT$. Given that $\bC$ has a simple analytical form in equation~\eqref{def_consistency}, so does its Jacobian, as detailed in~\S\ref{sec:ConsistencyJac}.

Equation~\eqref{iteration_Newton} is the second important difference between~\cite{Gruzinov+2020} and the present work.
Indeed, \cite{Gruzinov+2020} solves the self-consistency requirement on their equation~{(3)} by iteratively computing a sequence ${ \psi \!\to\! F \!\to\! \rho \!\to\! \psi \!\to\! \cdots }$, of potentials, \DFs\ and densities, up to convergence. Here, we directly ensure self-consistency in equation~\eqref{eq:Shape_from_Lagrange} by using Newton's method.

\subsection{Optimisation strategy}
\label{sub:OptStrat}

To finalise our algorithm, it only remains to specify our choice for the starting point $\bT_{0}$ of the iteration process.

In~\S\ref{sec:Random_Init} we explore a first method where $\bT_{0}$ is initialised at random. Such an agnostic approach allows us to recover a cluster's both stable, metastable and unstable equilibria, and to extend the results of~\cite{Takacs+2018} to (axisymmetric) multi-population clusters. Using this method, we recover all the qualitative behaviours of the single-population clusters reported in~\cite{Roupas+2017} and~\cite{Takacs+2018}, and confirm that the branch that goes through ${ \beta \!=\! 0 }$ (see Fig.~\ref{fig:Caloric_curves}) always has the highest entropy. Unfortunately, as one increases the number of stellar populations, the efficiency of this protocol drastically drops.

Dealing with systems with many stellar populations therefore requires improvements to the initialisation process. To alleviate most of these difficulties, we restrict ourselves, and predict only the clusters' global thermodynamical equilibria, and none of the other possible equilibria, should they be unstable or metastable. Benefiting from the insight of Fig.~\ref{fig:Caloric_curves}, we obtain these global equilibria by iteratively moving along the series of equilibria associated with the branch that has a solution for ${ \beta \!=\! 0 }$ (see~\S\ref{sec:Iterative_resolutions}). Of course, the main drawback of this approach is that it cannot be used to determine any of the unstable thermodynamical equilibria.

\section{Thermodynamical nuclear equilibria}
\label{sec:application}

We can now make use of our entropy optimisation algorithm to investigate the typical equilibrium distribution of orientations in galactic nuclei. Benefiting from the efficiency and versatility of this method, we will also use it to explore the (large) parameter space describing possible initial stellar clusters.

\subsection{Parametrisation of the stellar population}
\label{sub:param}

\begin{figure}
	\centering
	\includegraphics[width=0.38 \textwidth]{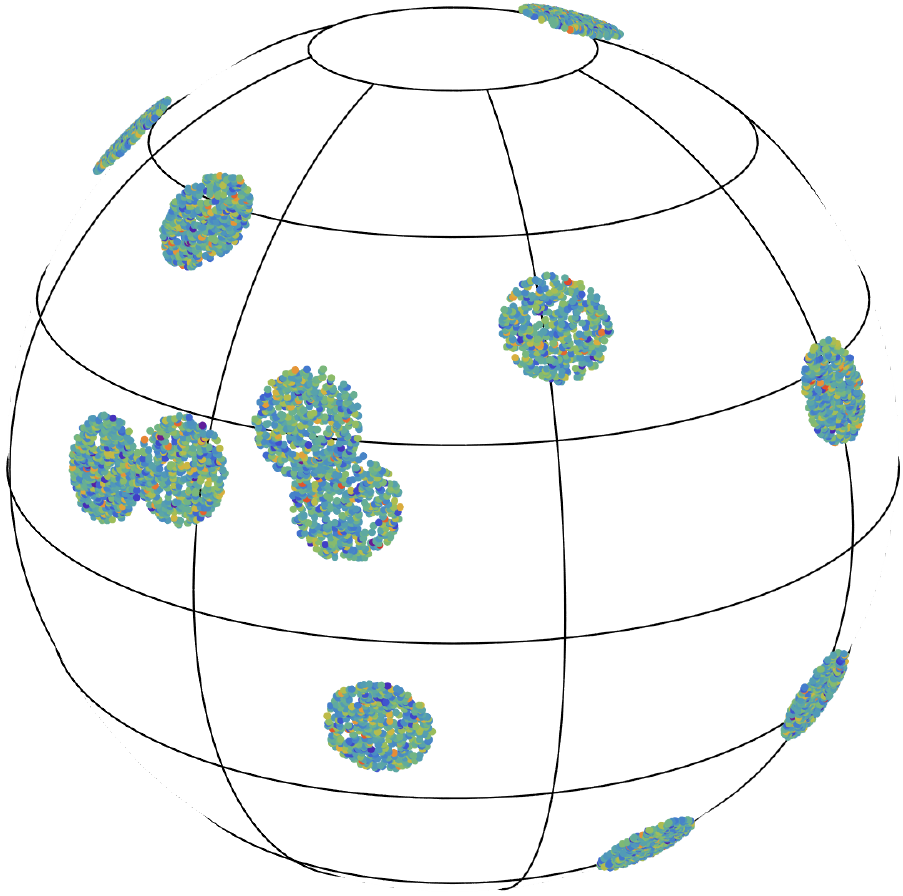}
	\caption{Illustration of a cluster's typical initial condition. It consists of ${ \Ndisc \!=\! 16 }$ dense patches of stars scattered uniformly on the unit sphere, and associated with distinct episodes of star formation or infall events. Each patch is made of ${ \Npart \!=\! 512 }$ stars distributed uniformly within a small angular section around the patch's centre. Each star is coloured according to the norm of its angular momentum vector, with the smallest norm in blue and the largest one in red. Clearly, stars with small angular momenta are the most numerous.}
	\label{fig:Initial_Discs}
\end{figure}

Let us first parametrise the distribution of the orbital parameters $\bK_{k}$, i.e., the distribution of masses, semi-major axes, and eccentricities. Our fiducial model is the same as in~\cite{Szolgyen+2018}. We assume that stars are formed through a series of distinct episodes of star formation or infall events. More precisely, for a given realisation, we generate ${ \Ndisc \!=\! 16 }$ discs, each of them composed of ${ \Npart \!=\! 512 }$ stars, so that the cluster's total number of particles is ${ N \!=\! \Ndisc \!\times\! \Npart \!=\! 8\,192}$. For each disc, its average orientation, $\hbLdisc$, is drawn uniformly on the the unit sphere, while the orientations of the stars within that disc are drawn uniformly within the small region ${ \hbL \!\cdot\! \hbLdisc \!\geq\! 0.994 }$. Finally, for each star, the orbital parameters ${ \bK \!=\! (m , a, e) }$ are drawn independently from one another according to power law \PDFs\ proportional to ${ (m^{-2}, a^{0},e) }$ respectively within the ranges ${ \mmax/\mmin \!=\! 100 }$, ${ \amax / \amin \!=\! 100 }$, and ${ (\emin , \emax) \!=\! (0,0.3) }$. In Fig.~\ref{fig:Initial_Discs}, we illustrate one typical realisation of such a protocol.
Following~\S\ref{sub:AxiAssump}, we recall
that the realisations are always rotated to have
their $\bLtot$ aligned along ${ +z }$.

\begin{figure}
	\centering
	\includegraphics[width=0.45 \textwidth]{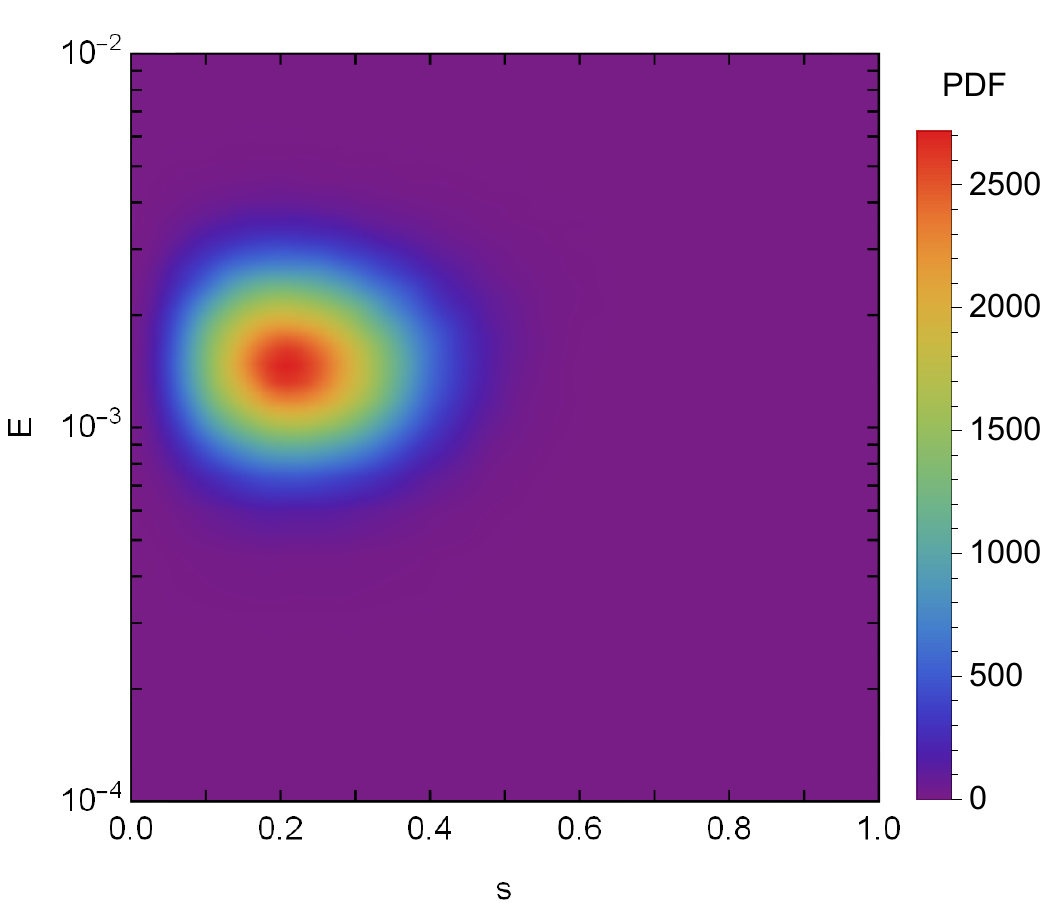}
	\caption{Distribution of the spin and binding energy of clusters drawn following the fiducial protocol from~\S\ref{sub:param}, for $2^{18}$ realisations of initial conditions and ${ \ellmax \!=\! 10 }$.}
	\label{fig:Histogram}
\end{figure}

Once an initial distribution has been drawn, one may compute its two key invariants, $\Etot$ and ${ |\bLtot| }$. In practice, we keep track of these two invariants through the two dimensionless quantities
\begin{equation}
	E = - \frac{\Etot}{N^{2} \, G \mmin^{2} / \amin} ;
	\quad
	s = \frac{|\bLtot|}{\sum_{i} L (\bK_{i})} ,
	\label{def_E_s}
\end{equation}
which we respectively call the binding energy and spin of the stellar cluster.
In Fig.~\ref{fig:Histogram}, we represent the typical distribution of ${ (E , s) }$ for a large number of stellar clusters drawn from our fiducial procedure.
We find that clusters exhibit a wide range of binding energy and spin, and in turn this somewhat impacts the diversity of thermodynamical equilibria.

\subsection{A typical equilibrium distribution}
\label{sec:Typical}

\begin{figure}
	\centering
	\includegraphics[width=0.4 \textwidth]{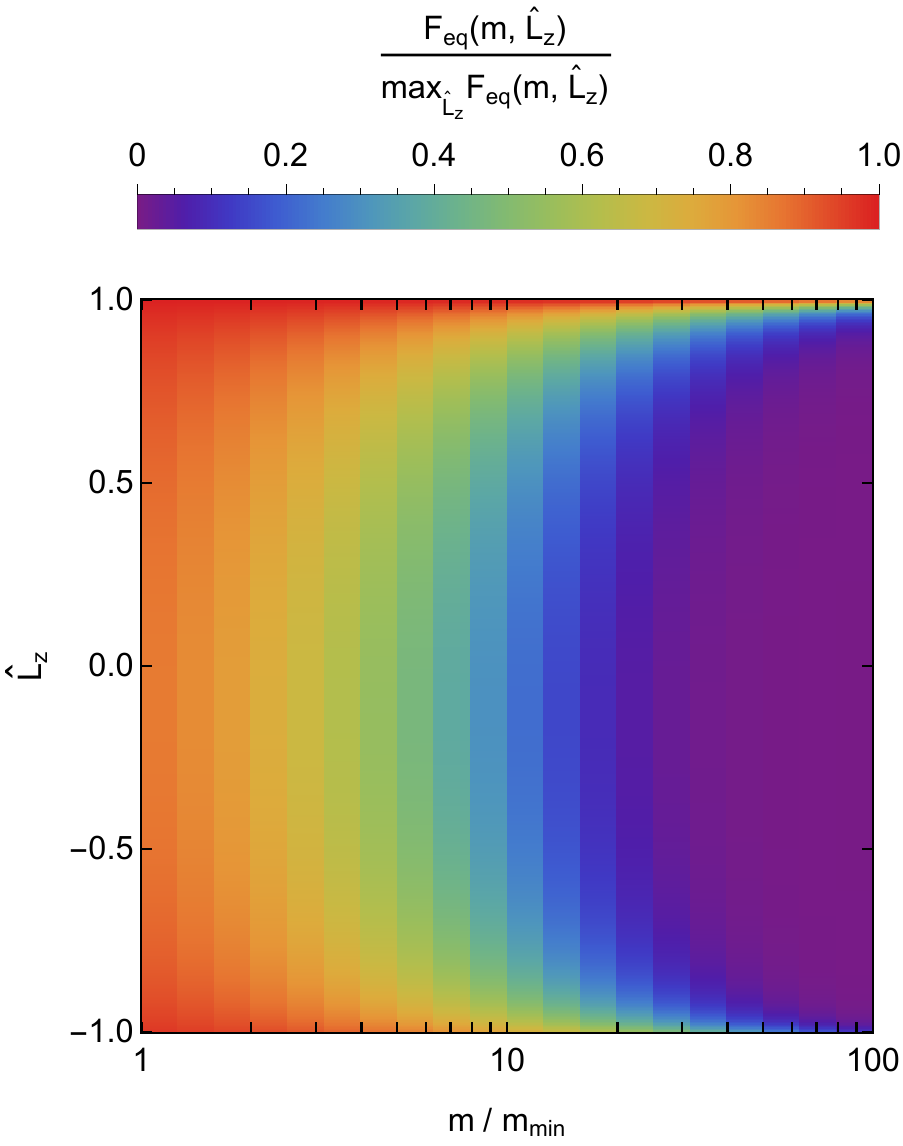}
	\includegraphics[width=0.5 \textwidth]{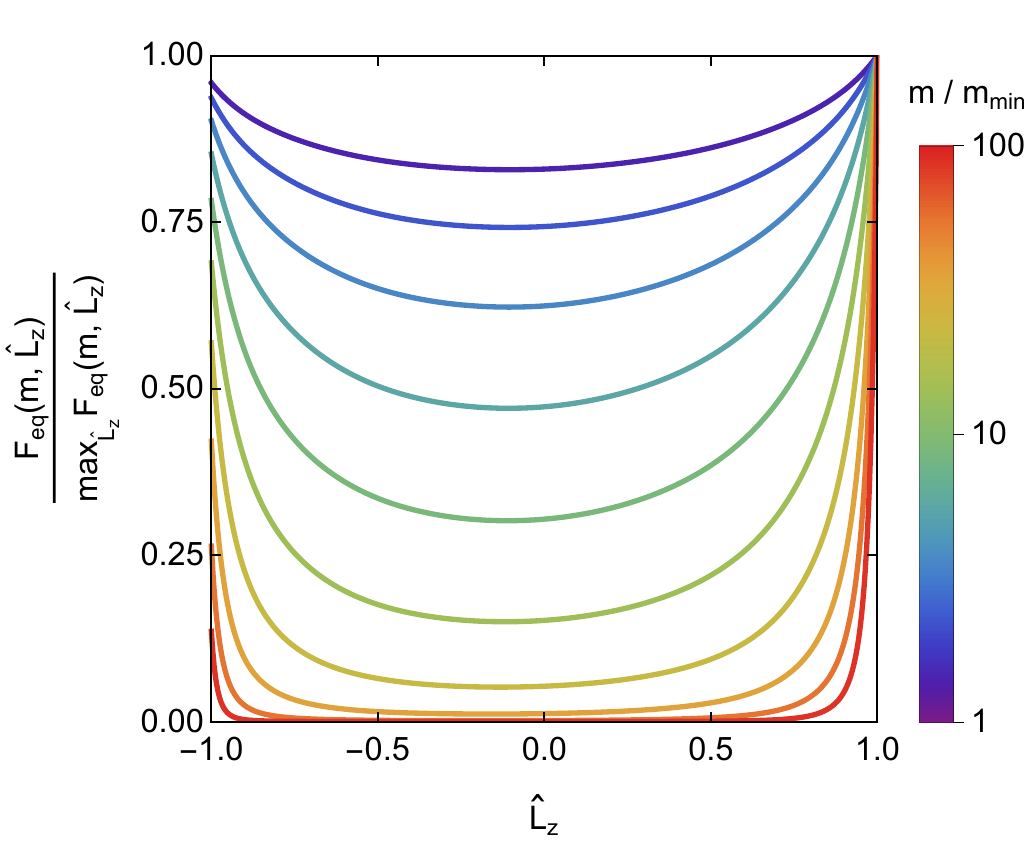}
	\caption{\textit{Top panel}: Illustration of the relaxed stellar density distribution, ${ \Feq (m , \hL_{z}) }$ (integrated over all semi-major axes and eccentricities), as a function of mass and orientation for a relaxed cluster with ${ (E , s) \!=\! (2 \!\times\! 10^{-3}, 0.2) }$. See~\S\ref{sec:NumericalDetails} for the numerical details. Our normalisation allows for each mass bin to have its maximum equal to 1. Heavier populations are much more segregated towards the poles.
	\textit{Bottom panel}: Same as the top panel, where each line corresponds to a different mass bin.}
	\label{fig:SegregationMass}
\end{figure}

\begin{figure}
	\begin{minipage}{0.22 \textwidth}
		\centering
		\includegraphics[width = \linewidth]{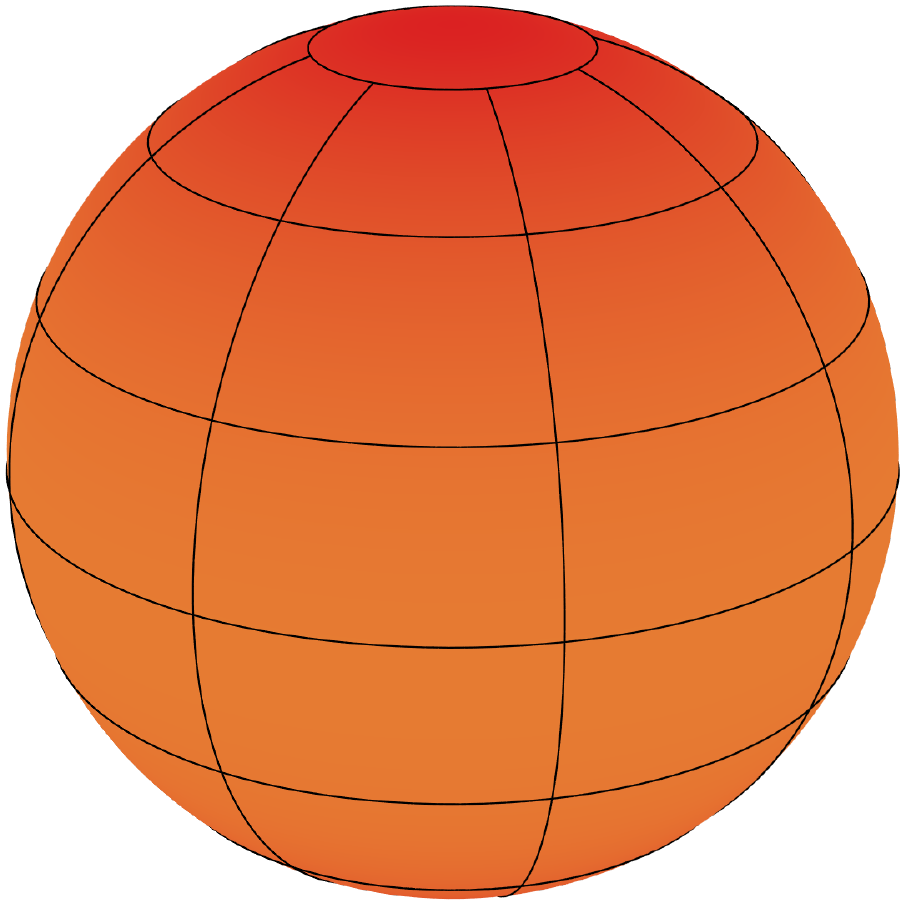}
	\end{minipage} \hfill
	\begin{minipage}{0.22 \textwidth}
		\centering
		\includegraphics[width = \linewidth]{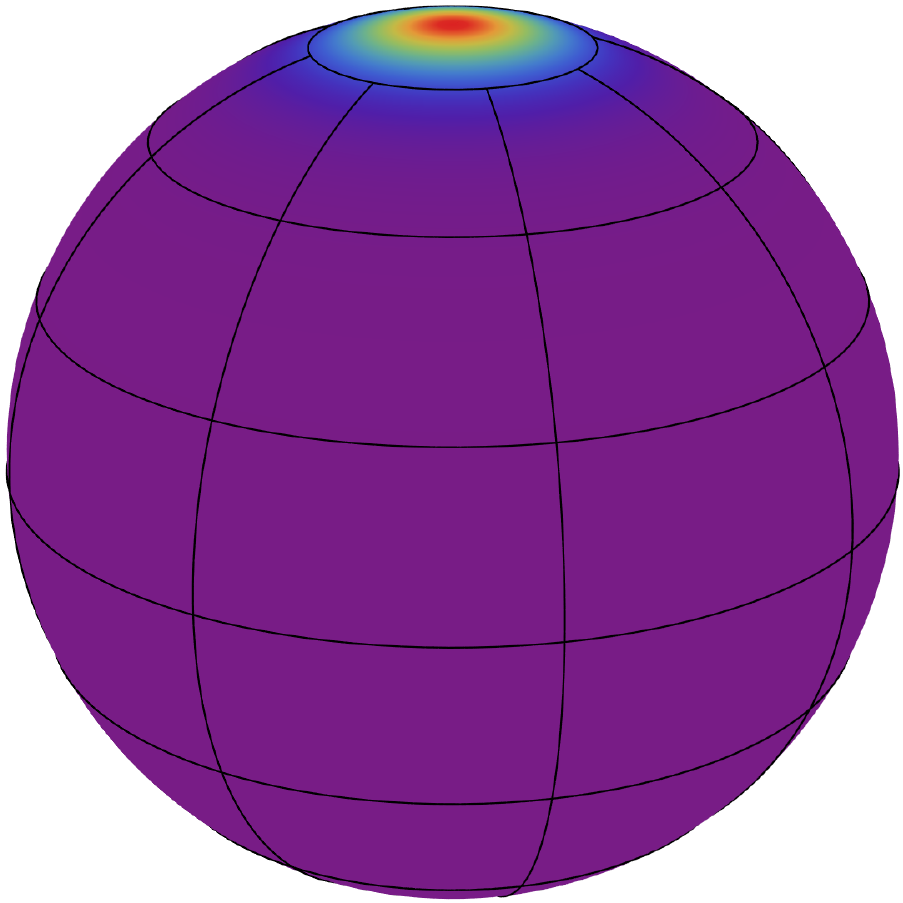}
	\end{minipage} \hfill
	\caption{Stellar density distribution on the unit sphere, after relaxation, for the lightest particles (left, ${ 1 \!\leq\! m/\mmin \!\leq\! 1.25 }$) and the heaviest ones (right, ${ 79 \!\leq\! m/\mmin \!\leq\! 100 }$). We use the same normalisation and colouring as in Fig.~\ref{fig:SegregationMass}. Importantly, we note that the heavy particles are significantly more segregated towards the poles, i.e.\ the corresponding heavy stars live within a thin equatorial disc.}
	\label{fig:Distribution_on_the_sphere}
\end{figure}

In Fig.~\ref{fig:Histogram}, we determined the distribution of the clusters' invariants ${ (E , s) }$. In order to characterise the properties of the associated systems, we now investigate in detail the particular case ${ (E ,s) \!=\! (2 \!\times\! 10^{-3} , 0.2) }$, i.e.\ a typical cluster realisation. We detail in~\S\ref{sec:DiscK} our precise choices for the discretisation of the stellar populations. By maximising the entropy, we determine this cluster's equilibrium configuration, ${ \Feq (\hbL , \bK) }$. This is first illustrated in Fig.~\ref{fig:SegregationMass} in the ${ (m , \hL_{z})}$-plane.
The same cluster is alternatively represented in Fig.~\ref{fig:Distribution_on_the_sphere}, where we present the stellar density distribution on the sphere for two different mass bins.

As is clearly visible, heavy particles tend to have their unit angular momentum vector $\hbL$ oriented towards the North pole, meaning that the associated stars tend to orbit near the equatorial plane, a conclusion already reached by~\cite{Szolgyen+2018} using an alternative Monte--Carlo approach\footnote{Note that even when accounting for the different normalisation, the details of the \DF\ in ${ (m,\hLz) }$-space obtained here in Fig.~\ref{fig:SegregationMass} somewhat differ from the equivalent fig.~{2} of~\cite{Szolgyen+2018}.}, and by~\cite{Gruzinov+2020} in the more general case of Keplerian elliptic wires. Having the heavy particles, e.g.\@, the \IMBHs, orbit close to the same orbital plane should drastically impact the rate of their pairwise mergers in galactic nuclei. In practice, we also repeated the experiment from Fig.~\ref{fig:SegregationMass} a hundred times by letting ${ (E,s) }$ explore the distribution from Fig.~\ref{fig:Histogram}. The associated ensemble-averaged \DF\ was found to be similar to the one from Fig.~\ref{fig:SegregationMass}.

For the heaviest stars, we also note the presence of an additional over-density near the South pole indicating the presence of a counter-rotating equatorial disc. However, owing to the conservation of angular momentum, this component is less populated than the main prograde disc. 

As one considers lighter particles, the anisotropy fades away and the lightest particles do not show any strong sign of spontaneous orientation alignment.
This concurs with \cite{Gruzinov+2020}'s results (see~\S{6.3.} therein),
which found that light objects follow a spherically symmetric distribution. 

\begin{figure}
	\centering
	\includegraphics[width=0.45 \textwidth]{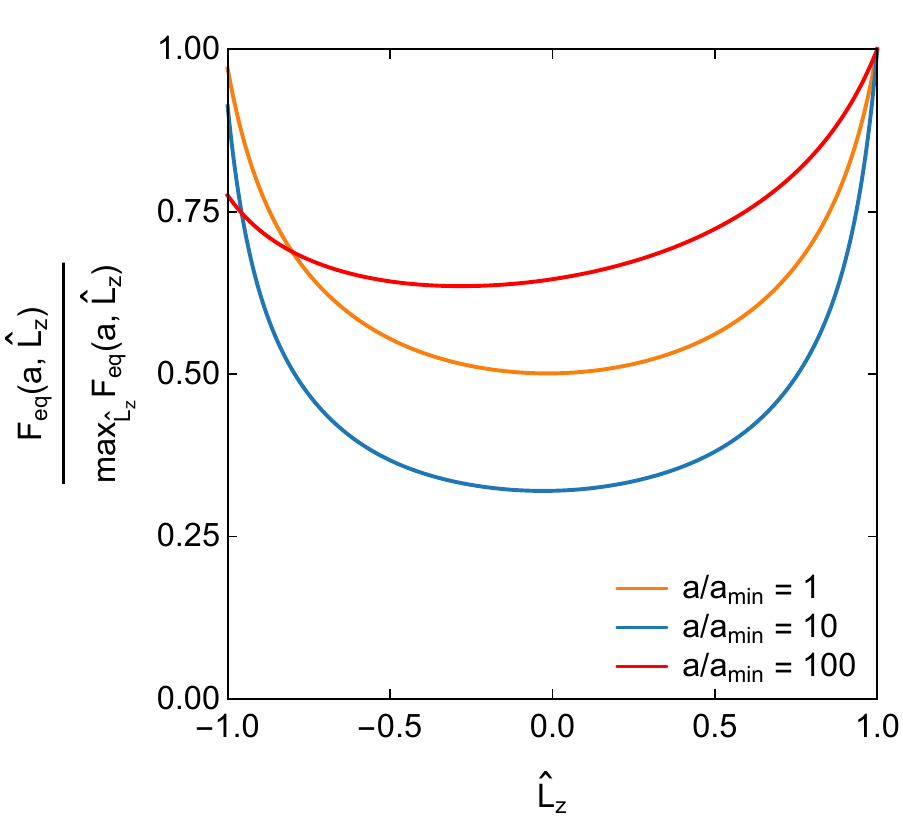}
	\caption{Illustration of the relaxed stellar density distribution, ${ \Feq (a , \hL_{z}) }$ (integrated over all eccentricities, but only intermediate masses in ${ 2 \!\leq\! m/\mmin \!\leq\! 16 }$), as a function of semi-major axis and orientation, for the same cluster as in Fig.~\ref{fig:SegregationMass}. The normalisation allows for each mass bin to have its maximum equal to 1. Populations with an intermediate semi-major axis tend to segregate within a thinner disk.}
	\label{fig:Effect_of_a}
\end{figure}

Having the full equilibrium distribution ${ \Feq (\hbL , \bK) }$ at our disposal, one may also study its dependence w.r.t.\ the semi-major axis and eccentricity. Limiting ourselves to particles of intermediate mass, we find that the disc is slightly thinner for intermediate semi-major axes, as illustrated in Fig.~\ref{fig:Effect_of_a}.
This is in agreement with the recent result from~\cite{Mathe+2022} (see fig.\@~{7} therein).

As for eccentricities,
we find that, in essence, $\Feq$ is independent of $e$.
This was expected given that the coupling coefficients, ${ \mH_{\ell} [\bK , \bKp] }$ (equation~\ref{def:J_l}) and the norm of the angular momentum ${ L (\bK) }$ (equation~\ref{def_LK}) only weakly depend on $e$, especially for the chosen quasi-circular orbits ${ 0 \!\leq\! e \!\leq\! 0.3 }$.

\subsection{Impact of the cluster's binding energy and spin}

\begin{figure}
	\centering
	\includegraphics[width=0.45 \textwidth]{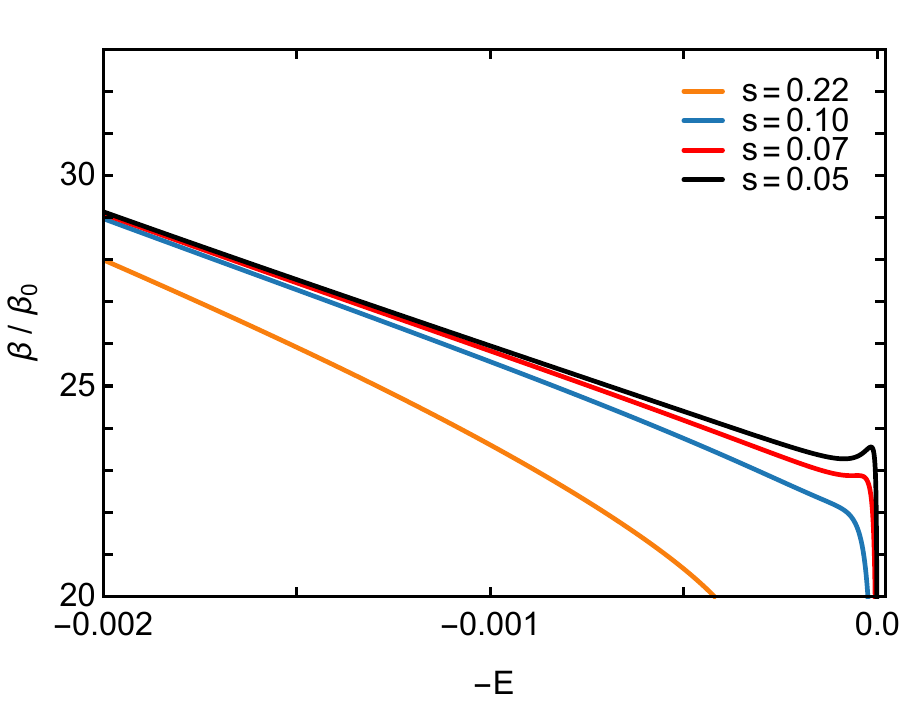}
	\includegraphics[width=0.45 \textwidth]{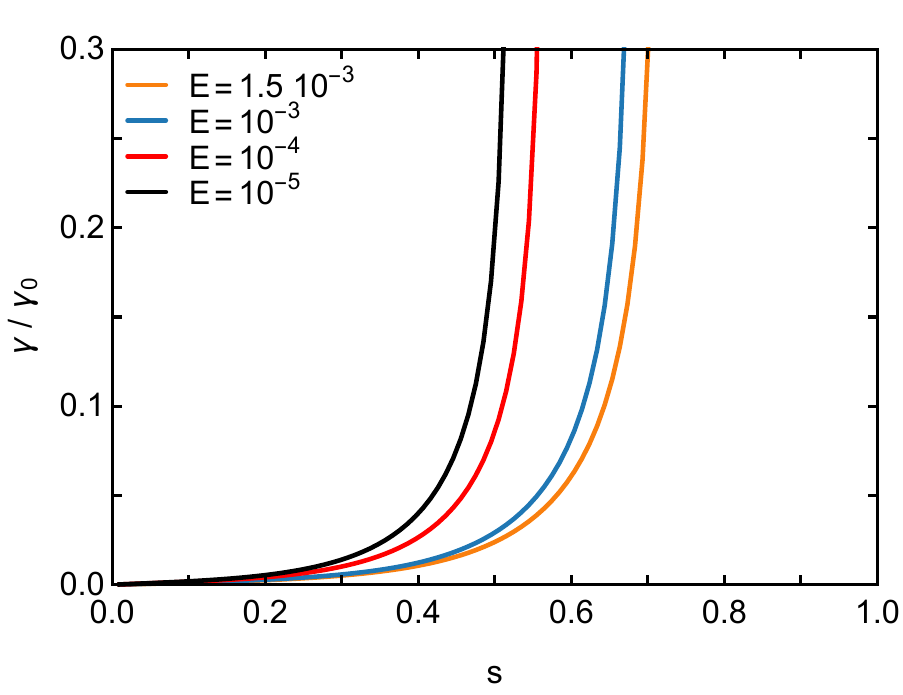}
	\caption{\textit{Top panel}: Illustration of the equilibrium inverse temperature $\beta$ as a function of the cluster's normalised total energy, ${ -E }$ (equation~\ref{def_E_s}) for different values of spin.
	Each line is composed of 500 systems,
	all with ${ (m, a, e) }$ drawn following the protocol of \S\ref{sub:param},
	and with binding energies distributed on a logarithmic grid ranging from ${ 10^{-6} }$ to ${ 10^{-1} }$.
	\textit{Bottom panel}: Illustration of the equilibrium angular frequency $\gamma$ as a function of the cluster's spin for different binding energies. For each line, 100 systems were simulated, with spins distributed on a linear grid ranging from 0 to 1.
	See equation~\eqref{def_range_random} for the definitions of $\beta_{0}$ and $\gamma_{0}$. As one lowers the total angular momentum, via $s$, thermodynamical equilibria can exhibit negative specific heats, i.e.\ ${ C \!=\! \p \Etot / \p T \!<\! 0 }$~\citep{Roupas+2017}. For a given value of $s$, our approach to determine the series of equilibria is to start from the (much) simpler problem at ${ \beta \!=\! 0 }$ and move up to the target energy step by step iteratively (see~\S\ref{sec:Iterative_resolutions}).}
	\label{fig:Series_of_energies}
\end{figure}

As highlighted in Fig.~\ref{fig:Histogram}, the clusters exhibit a significant diversity in their binding energies and spins, given our generation protocol. Let us therefore investigate the dependence of the equilibria w.r.t.\ these invariants. In Fig.~\ref{fig:Series_of_energies}, we present series of equilibria (caloric curves) giving the inverse temperature, $\beta$, as a function of the normalised total energy, ${ -E }$, for various total angular momentum (i.e.\ various $s$).
We observe that the temperature generically increases with energy, except in a small region of negative specific heat. Also, for any value of the spin there exists a region of negative temperature. Both effects had already been reported in~\cite{Roupas+2017} and~\cite{Takacs+2018}, and are further discussed in~\S\ref{sec:Random_Init}. Note however that these behaviours arise for ${ E \lesssim 10^{-4} }$, i.e. well outside of the astrophysical regime,
${ E \!\gtrsim\! 10^{-3} }$, found in Fig.~\ref{fig:Histogram}.
In the same figure, we also present series of equilibria giving the angular frequency $\gamma$ as a function of the total angular momentum $s$, for fixed values of the energy $E$. The angular frequency is found to always increase with the cluster's spin.

Within the domain of invariants spanned by Fig.~\ref{fig:Histogram}, all equilibria are found to remain qualitatively similar but differ in the strength of their anisotropy. To characterise the level of anisotropy in the orbital distribution, we introduce a cluster's segregation rate, $\SR$, as
\begin{equation}
	\SR = \frac{\displaystyle \int_{m \geq \mc} \!\!\!\!\!\!\!\!\!\!\!\!\!\! \rd \bK \;\; \int_{\hL_{z} \geq \hLc} \!\!\!\!\!\!\!\!\!\!\!\!\!\! \rd \hbL \, \Feq (\hbL , \bK)}{\displaystyle \int_{m \geq \mc} \!\!\!\!\!\!\!\!\!\!\!\!\!\! \rd \bK \;\; \int \!\! \rd \hbL \, \Feq (\hbL , \bK)} ,
	\label{def_SR}
\end{equation}
This quantity describes which fraction of heavy particles (i.e.\ with ${ m \!\geq\! \mc }$) have an angular momentum vector that is well-aligned with the cluster's total angular momentum (i.e.\ have ${ \hL_{z} \!\geq\! \hLc }$). With such a definition, we note that anti-aligned particles are not accounted for. The larger $\SR$, the stronger the anisotropic segregation of the heavy populations. In practice, guided by Fig.~\ref{fig:SegregationMass}, we consider the values ${ \mc \!=\! 10 \, \mmin }$ and ${ \hLc \!=\! \cos (20^\circ) }$. We expect that changing the values of $\mc$ does not qualitatively impact the present conclusions.

\begin{figure}
	\centering
	\includegraphics[width=0.45 \textwidth]{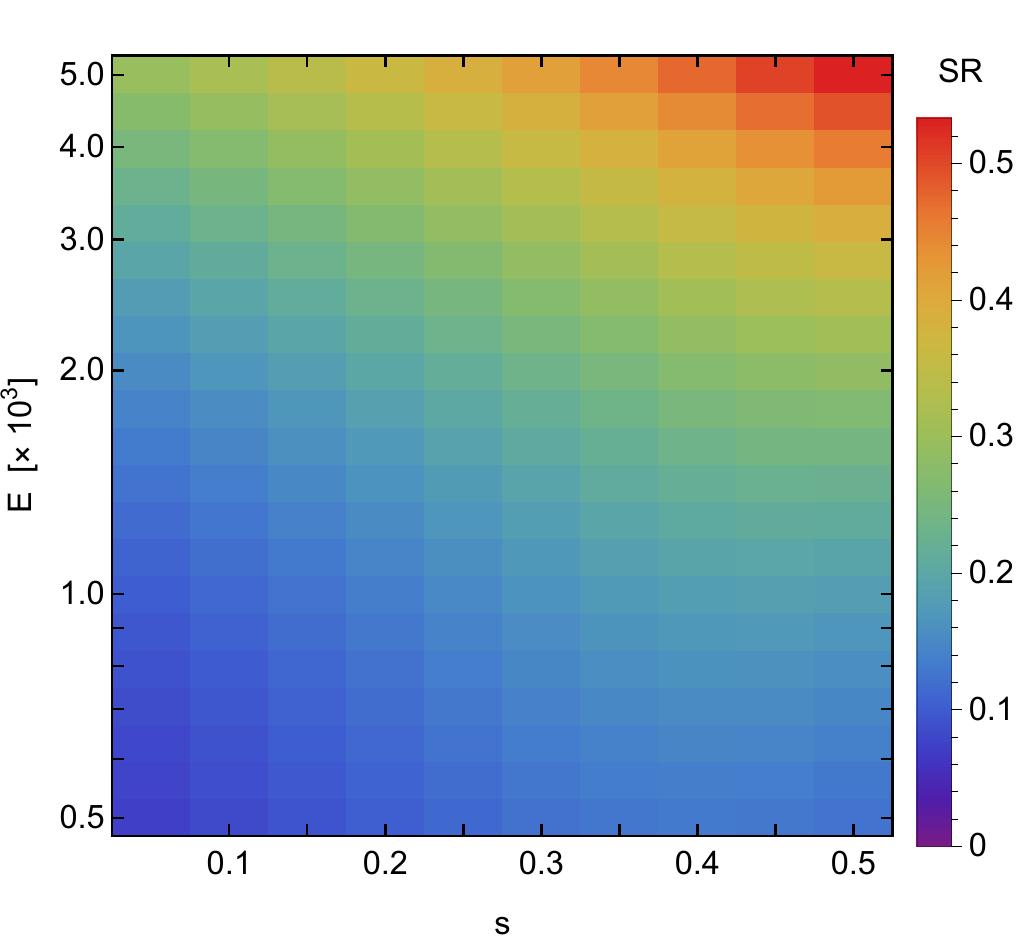}
\includegraphics[width=0.45 \textwidth]{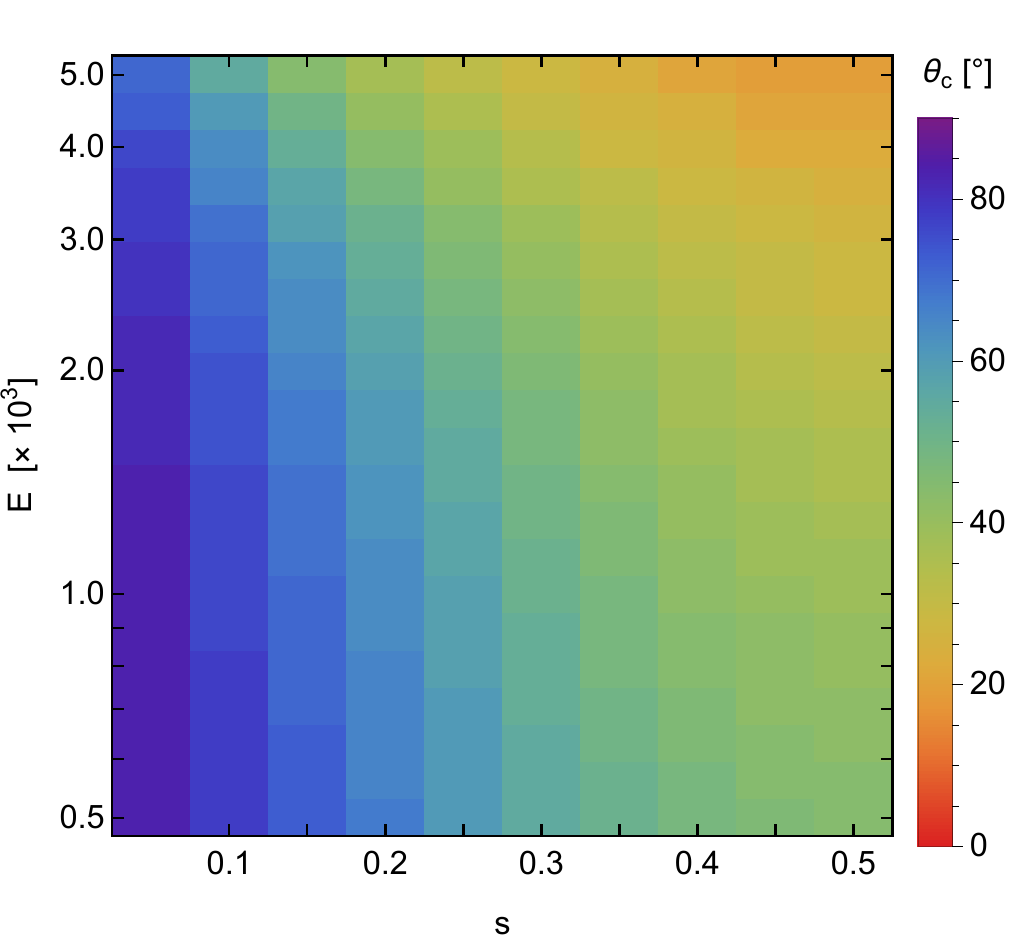}
	\caption{\textit{Top panel}: Segregation rate of the relaxed clusters (equation~\ref{def_SR}), within the range of spins and binding energies expected for typical astrophysical clusters (see Fig.~\ref{fig:Histogram}).
	\textit{Bottom panel}: Same domain as the top panel but for the segregation angle $\thetac$ (equation~\ref{def_thetac}). Clusters with large binding energies and spins exhibit the strongest segregation of their heavy particles within the equatorial disc. }
	\label{fig:Segregation}
\end{figure}

In Fig.~\ref{fig:Segregation}, we illustrate the clusters' segregation rate as a function of the binding energy and spin ${ (E,s) }$.
We observe the crucial role played by these two invariants in driving the spontaneous alignment of the orbital orientations of the heaviest particles. The larger $E$ and the larger $s$, the larger the proportion of heavy particles that are aligned near the equatorial plane.

To strengthen this conclusion, let us finally determine the angular size, $\thetac$, of the northern polar cap that contains a fraction $\fc$ of the heavy particles' unit angular momentum vectors $\hbL$. More precisely, we define $\thetac$ through the implicit constraint
\begin{equation}
	\frac{\displaystyle \int_{m \geq \mc} \!\!\!\!\!\!\!\!\!\!\!\!\!\! \rd \bK \;\; \int_{\hL_{z} \geq \cos (\thetac)} \!\!\!\!\!\!\!\!\!\!\!\!\!\! \rd \hbL \, \Feq (\hbL , \bK)}{\displaystyle \int_{m \geq \mc} \!\!\!\!\!\!\!\!\!\!\!\!\!\! \rd \bK \;\; \int \!\! \rd \hbL \, \Feq (\hbL , \bK)} = \fc ,
	\label{def_thetac}
\end{equation}
noting once again that anti-aligned stars are not accounted for.
Equivalently, $\thetac$ is also the angular size of the equatorial disk that contains a fraction $\fc$ of the heavy particles. Using a fraction ${ \fc \!=\! 50 \% }$, the dependence of $\thetac$ w.r.t.\ the clusters' invariants is illustrated in Fig.~\ref{fig:Segregation}. Once again, the more bound the cluster, and the larger its total angular momentum, the thinner the disc of heavy particles in the relaxed equilibrium.

Interestingly, if equations~\eqref{def_SR} and~\eqref{def_thetac} were to consider both aligned and anti-aligned stars, through the replacement ${ \hL_{z} \!\geq\! \cos (\thetac) \!\to\! |\hL_{z}| \!\geq\! \cos (\thetac) }$, the dependence w.r.t.\ $s$ in Fig.~\ref{fig:Segregation} would be significantly reduced. This result is expected since in equation~\eqref{def_Etot}, the total energy is left invariant by the changes ${ \hbL_{i} \!\to\! - \hbL_{i} }$, while in equation~\eqref{def_Ltot} the total angular momentum changes sign.

\subsection{Impact of the mass and semi-major axes distributions}
\label{sec:Impact_m_a}

Let us now step out and explore the impact of the distribution of the orbital stellar parameters themselves on the clusters' thermal equilibria. Following~\S\ref{sub:param}, we now vary the \PDFs\ w.r.t.\ which the stars' masses and semi-major axes are drawn, keeping all other parameters the same. More precisely, we still draw ${ (m,a) }$ pairs according to power law distributions, ${ (m^{\gamma_{m}} , a^{\gamma_{a}} ) }$, but this time vary the power law indices ${ (\gamma_{m} , \gamma_{a}) }$ between different realisations. 

\begin{figure}
	\centering
	\includegraphics[width=0.47 \textwidth]{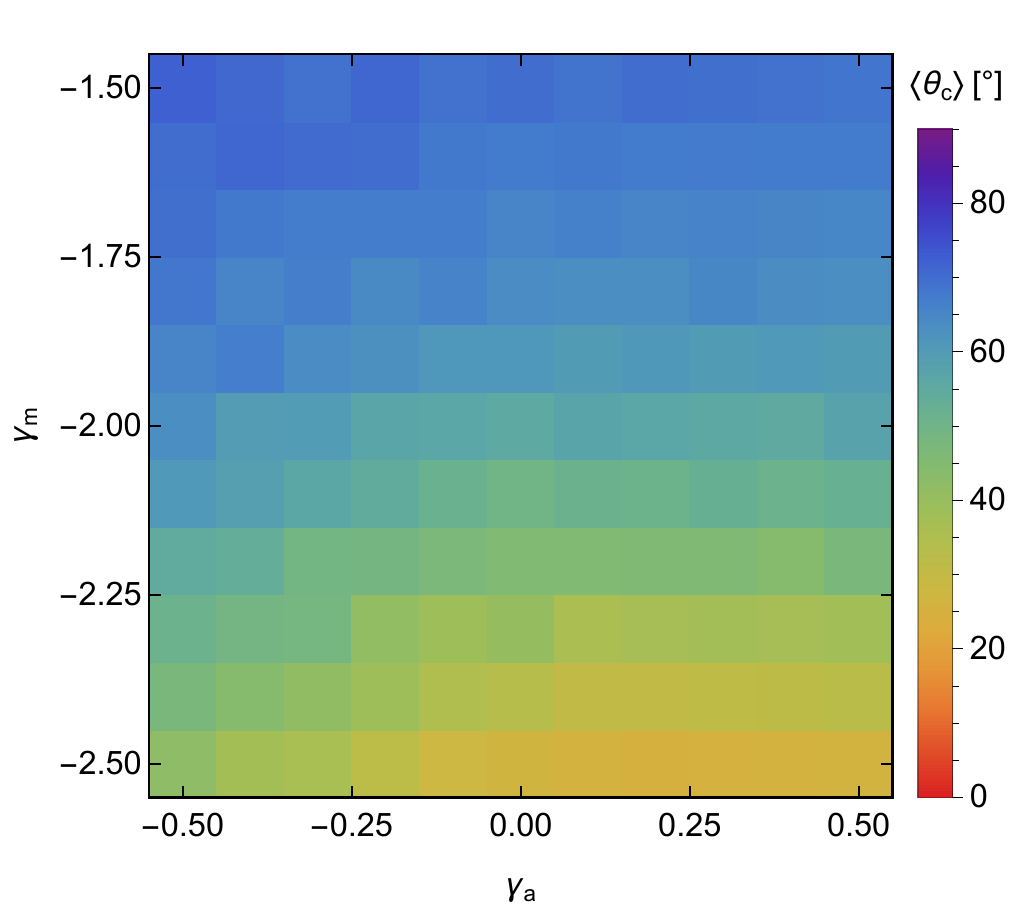}
	\caption{Average segregation angle, ${ \langle \thetac \rangle }$, as a function of the power law indices of the initial distributions in semi-major axes and masses. Here, for each value of ${ (\gamma_{a} , \gamma_{m}) }$, the average has been computed over 128 realisations of a cluster. The fiducial case from Fig.~\protect\ref{fig:SegregationMass} corresponds to ${ (\gamma_{a} , \gamma_{m}) \!=\! (0,-2) }$. Interestingly, both power law indices are found to impact the relaxed segregation angle.
	}
	\label{fig:Effect_Gamma}
\end{figure}

The impact of changing these two parameters on the average segregation rate of the heavy particles, ${ \langle \thetac \rangle }$, is illustrated in Fig.~\ref{fig:Effect_Gamma}.
We find that the slope $\gamma_{m}$ has the strongest effect with clusters containing fewer heavy stars segregating in a disk twice thinner than the fiducial cluster. The semi-major axis power law index also has an impact on the segregation angle: clusters with larger semi-major axes tend to display a stronger mass segregation. 

These dependencies may possibly be used in the future to estimate indirectly the stellar and compact objects' initial mass function in galactic centres. Indeed, the rate of pairwise IMBH mergers should be measurable using gravitational waves, and could serve as an indirect probe for the width of the segregated massive disc, which we just have shown is linked to the \IMBHs\@' initial mass function.

\subsection{Impact of the dispersion of disc orientations}

A cluster's long-term equilibrium distribution is fully characterised by its three invariants: the binding energy, $E$, the spin, $s$, and the distribution of orbital parameters, ${ N (\bK) }$. Unfortunately, while dynamically relevant, such parameters do not translate easily in terms of astrophysical observables. As such, let us finally switch to parametrisations more closely related to the underlying formation process of the galactic centre's stellar cluster, such as the geometry of the initial distribution.

\begin{figure}
	\centering
	\includegraphics[width=0.38 \textwidth]{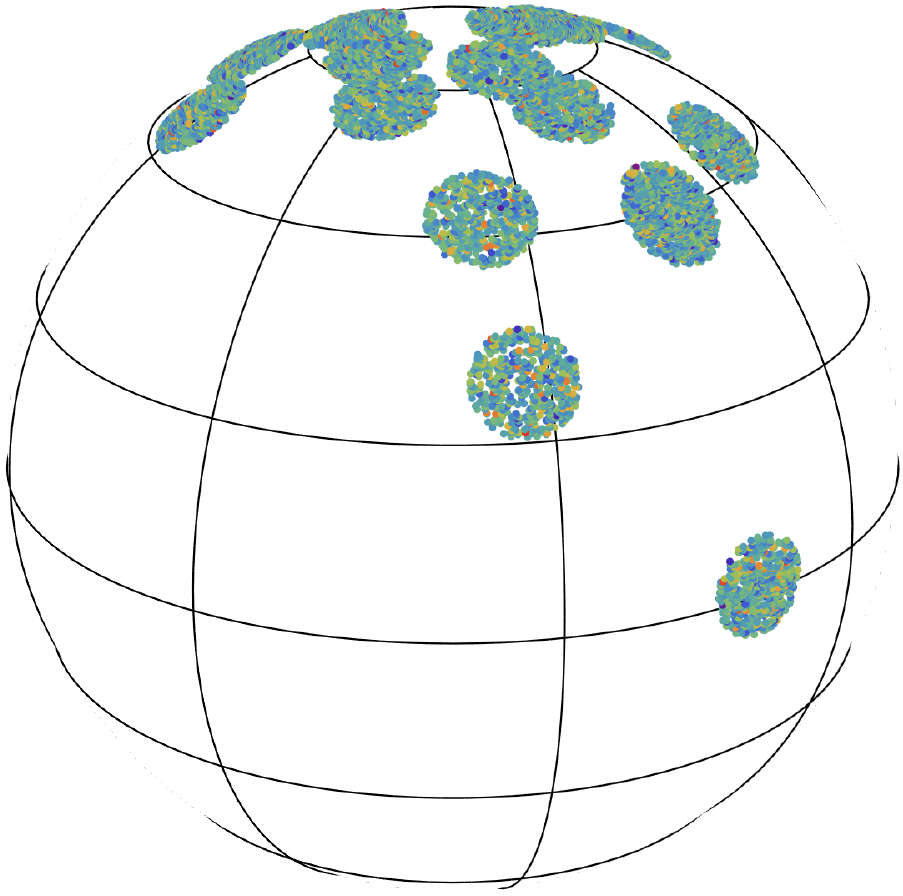}
	\caption{Same as Fig.~\ref{fig:Initial_Discs}, but assuming that the orientations of the discs are drawn from a von Mises--Fisher \PDF\ with concentration ${ \kappa \!=\! 5 }$ (equation~\ref{def_vonMises}). }
	\label{fig:Narrow_Discs}
\end{figure}

Building upon Fig.~\ref{fig:Initial_Discs}, rather than drawing the initial orientations of the discs uniformly on the unit sphere, we may draw them along some biased direction, for example to reflect the preferential infall of new star-forming gas along specific directions, e.g.\@, imposed by the geometry of past gaseous accretion events. More precisely, we assume that the average orientation of each disc is drawn according to a Von Mises--Fisher \PDF\@~\citep{Wood1994} of the form
\begin{equation}
	P (\hbL) = \frac{\kappa}{4 \pi \sinh (\kappa)} \, \re^{\kappa \hL_{z}} ,
	\label{def_vonMises}
\end{equation}
with $\kappa$ the \PDF\@'s concentration. Here, ${ \kappa \!=\! 0 }$ corresponds to the isotropic case considered in Fig.~\ref{fig:Initial_Discs}. The larger $\kappa$, the smaller the spread of the \PDF\ on the unit sphere, as in Fig.~\ref{fig:Narrow_Discs} for ${ \kappa \!=\! 5 }$.

\begin{figure}
	\centering
	\includegraphics[width=0.45 \textwidth]{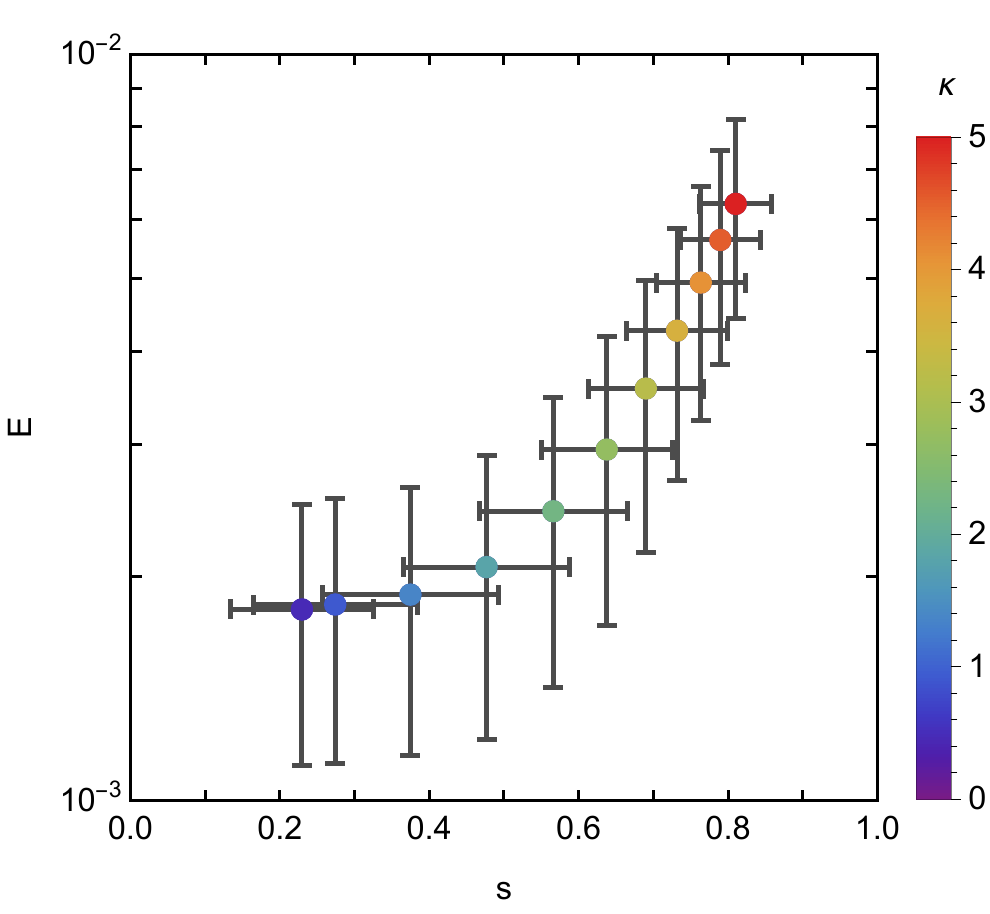}
	\caption{Typical distribution of the spin and binding energies as one varies the concentration $\kappa$ of the initial discs. Coloured points correspond to the mean values of ${ (s,E) }$, while black error bars are standard deviations in both parameters. Statistics are computed from $10^{4}$ realisations. As expected, the more concentrated the discs' orientations, the more packed the initial distributions, and therefore the larger the spins and binding energies. }
	\label{fig:Histogram_Narrow}
\end{figure}

\begin{figure}
	\centering
	\includegraphics[width=0.45 \textwidth]{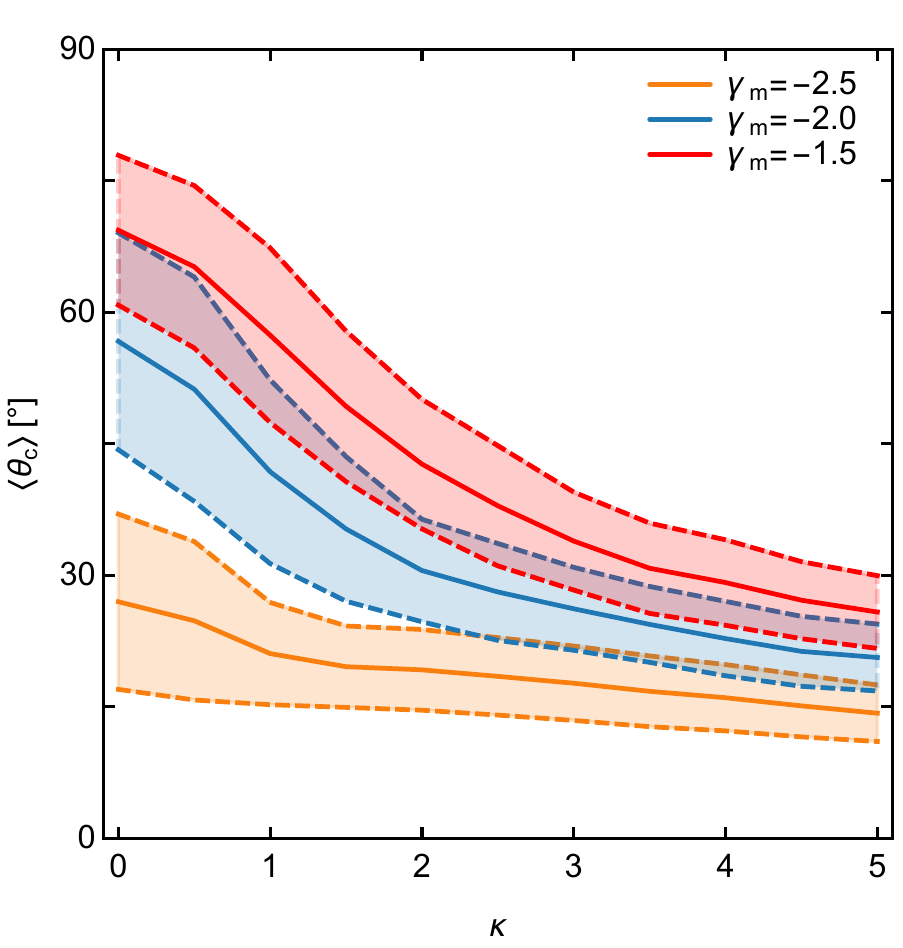}
	\caption{Average segregation angle, ${ \langle \thetac \rangle }$, as a function of the concentration parameter $\kappa$, for three different mass distributions. The solid lines represent the average segregation angle, and the dashed lines represent the standard deviation. The statistics are estimated from $2^{10}$ realisations -- though ${ \sim\! 10\% }$ of the realisations for ${ (\kappa,\gamma_{m}) \!=\! (5,-2.5) }$ failed because of numerical overflows encountered with very strongly anisotropic \DFs\@. The more anisotropic a cluster initially, the higher its binding energy and spin, and therefore the more segregated at equilibrium.}
	\label{fig:SA_Narrow}
\end{figure}

Keeping all other stellar parameters as in~\S\ref{sub:param}, we may now study the impact of the concentration parameter $\kappa$ on the shape of the relaxed distributions. This is first illustrated in Fig.~\ref{fig:Histogram_Narrow}, where we show the dependence of the invariants ${ (E,s) }$ with $\kappa$.
As expected, we recover that the more concentrated the distribution of the discs' initial orientations, the larger the value of $s$ and $E$. In Fig.~\ref{fig:SA_Narrow}, we subsequently illustrate the dependence of the average segregation angle, ${ \langle \thetac \rangle }$ (equation~\ref{def_thetac}), as a function of $\kappa$.
Narrower initial distributions of orientations lead to stronger segregation of the heavy stars at equilibrium. This is expected, since Fig.~\ref{fig:Segregation} showed that the segregation strength correlates positively with both binding energy and spin. To emphasise this conclusion, Fig.~\ref{fig:SA_Narrow} also illustrates the thickness of the asymptotic massive disc as a function of the initial mass function slope. Similarly to Fig.~\ref{fig:Effect_Gamma}, at fixed initial anisotropy, the steeper the slope, the stronger the mass segregation. Overall, investigations as in Fig.~\ref{fig:SA_Narrow} should ultimately prove useful to place some constraints on the origin of SgrA*'s surrounding stellar cluster,
in particular in the light of the ``paradox of youth''~\citep{Ghez+2003,Genzel+2010}.

\section{Conclusion}
\label{sec:conclusion}

In the spirit of~\cite{Roupas+2017,Takacs+2018}, we relied on maximum entropy methods to assess the internal stellar structure of galactic nuclei and their underlying distribution of orbital orientations. We expanded their approach to multi-populations clusters and showed how to jointly account for the constraints of energy, angular momentum, and orbital parameters conservations to efficiently characterise a cluster's expected thermodynamical equilibria. In particular, as already pointed out in~\cite{Szolgyen+2018}, we recovered the spontaneous alignment of the heavier stellar components, such as \IMBHs\ (Fig.~\ref{fig:SegregationMass}). The more bound the cluster and the larger its spin, the thinner the asymptotic disc (Fig.~\ref{fig:Segregation}).

Benefiting from the versatility of the present method, we used it to explore the dependence of these structures w.r.t.\ the cluster's initial conditions of formation, such as the intrinsic power law spectra in mass and semi-major axis (Fig.~\ref{fig:Effect_Gamma}), or the initial spread in orientations (Fig.~\ref{fig:SA_Narrow}). Small modifications to the present algorithm would also make it possible to confirm that the final equilibrium does not depend on the specific sequence of star formation events~\citep{Szolgyen+2018}.

Of course, this work is but one step towards the systematic exploration of the remaining imprints of a galactic nucleus' initial formation on the cluster's long-term distribution of orientations. Let us conclude by listing a few possible venues for future works.

The present approach could first benefit from various extensions on the theoretical and analytical front. (i) We have assumed from the start that stellar eccentricities were conserved throughout the \VRR\ evolution. This amounts to neglecting the process of \SRR\@~\citep{Rauch+1996}, whose signature is clearly visible in the S-cluster distribution~\citep[see, e.g.\@,][]{Tep+2021}. As already recently hinted in~\cite{Szolgyen+2021}, allowing for the eccentricities to also vary during the orientations' relaxation might impact the system's \VRR\ equilibria. This deserves careful study. Similarly, stars can also change their semi-major axis through \NR\@, whose impact on the outcome of \VRR\ should also be investigated. (ii) We recovered here that \VRR\ may lead to highly non-spherical distribution of orientations. Such an end-state distribution is expected to impact in turn the rate of orientation diffusion of a given test star and the efficiency with which newly formed stellar discs can dilute~\citep[see, e.g.\@,][]{Giral+2020}.

The stochastic process of spontaneous alignment is not limited to the axisymmetric relaxation of Keplerian annuli around supermassive \BHs\@. As such, at a significantly larger numerical cost, one could extend the present approach to non-axisymmetric distributions, as was already done in~\cite{Roupas+2017} in the limit of a single-population system with a quadrupolar interaction. Similarly, it would be of interest to investigate the distinction between the present global equilibria and the other possible metastable ones\footnote{For systems with long-range interactions, metastable states have a very long lifetime scaling as ${ \re^{N} }$~\citep[see, e.g.\@,][]{Chavanis+2006}, because a system trapped in a metastable state (local entropy maximum) has to cross a huge barrier of potential to reach the fully stable equilibrium  (global entropy maximum).}: do they also exhibit a disc-like structure? Finally, one should investigate whether there exist astrophysically relevant regimes with negative temperatures or heat capacities (Figs.~\ref{fig:Series_of_energies} and~\ref{fig:Caloric_curves}).

In equation~\eqref{def:Htot}, we assumed that we could average the cluster's Hamiltonian over the in-plane precessions. We could lift this assumption and determine the equilibrium distributions of Keplerian elliptic wires~\citep[see, e.g.\@,][]{Gruzinov+2020,Tremaine2020III}. Finally, since the mean potential remains on average spherically symmetric, all these investigations could also be performed in globular clusters, where stellar metallicities or ages could be used as additional tracers of the mixing of stellar orientations.

In the present work, we restricted ourselves to predicting the end-state of \VRR\@, i.e.\ its thermodynamical equilibrium. By design such an approach cannot provide any estimate of the expected relaxation time required for this asymptotic distribution to be reached. There are, at least, two possible venues to quantify such relaxation time: (i) One could use direct time integrations of the equations of motion to get the equilibrium distributions, a goal already pursued in~\cite{Kocsis+2015}. Building upon~\cite{Fouvry+2020}, one can expect that efficient multipole methods may be designed to perform such direct numerical simulations with a computational complexity scaling linearly with the total number of particles; (ii) In the limit of sufficiently symmetric orbital distributions, e.g.\@, axisymmetric~\citep{Fouvry+2019Axi}, one could alternatively derive an explicit kinetic theory for the cluster, from which the relaxation time would naturally follow. Interestingly, we report that the axisymmetric equilibrium recovered in Fig.~\ref{fig:SegregationMass} generically exhibits a monotonic profile of latitudinal precession frequency for all stellar populations. This may induce a situation of ``kinetic blocking'' and play a critical role in defining the efficiency with which these systems may relax~\citep{Fouvry+2019Axi}.

Finally, benefiting from the planned upgrade on VLTI~\citep{Eisenhauer2019,Gravity+2021}, as well as the future thirty-meter class telescopes such as ELT~\citep{Pott+2018,Davies+2018} and TMT~\citep{Do+2019}, we will soon have a wealth of orbitally-resolved stars around SgrA* along with their stellar ages. Building upon the present work, a detailed characterisation of their orbital distribution should prove paramount to place constraints on the properties of the (likely present) un-observed \IMBHs\@, and the impact of their effective distribution of orientations on their overall in-situ merger rates.

\section*{Acknowledgements}

This work is partially supported by the grant \href{http://www.secular-evolution.org}{Segal} ANR-19-CE31-0017 of the French Agence Nationale de la Recherche, and by the Idex Sorbonne Universit\'e. NM acknowledges support from SUPAERO, the Universit\'e Paul Sabatier and the \'Ecole polytechnique. We thank St\'ephane Rouberol for running smoothly the Infinity cluster, where the simulations were performed.

\section*{Data availability}

The data and numerical codes underlying this article were produced by the authors. They will be shared on reasonable request to the corresponding author. The code is distributed on Github at the following URL: \href{https://github.com/NathanMagnan/vrroomE}{https://github.com/NathanMagnan/VRROOMe}.

\appendix

\section{Thermodynamical equilibrium}
\label{sec:Feq}

Following~\cite{Roupas+2017}, we derive the generic expression of the \DF\@ of thermodynamical equilibria, $\Feq$, from equation~\eqref{eq:Shape_from_Lagrange}. This \DF\ maximises the Boltzmann entropy (equation~\ref{def_S}), under three joint constraints: the conservation of orbital parameters (equation~\ref{def_NK}), the conservation of energy (equation~\ref{def_Etot}) and the conservation of angular momentum (equation~\ref{def_Ltot}).

Following the double orbit-average from equation~\eqref{def:Htot}, we stress that the Hamiltonian of \VRR\ (equation~\ref{eq:Htot}) is such that (i) the phase space domain for $\hbL$ is of finite volume and (ii) for a finite $\ellmax$, there is no divergence of the pairwise \VRR\ interaction for ${ \hbL_{i} \!=\! \hbL_{j} }$. As such, the \VRR\ dynamics is simplified and does not present the typical peculiarities of the statistical mechanics of self-gravitating systems such as the evaporation of stars, the formation of binaries, the gravothermal instability, and the absence of a true statistical equilibrium~\citep[see, e.g.\@,][]{Chavanis+2006}. In the present case, we essentially deal with the dynamics of self-interacting spins on a sphere, a system which possesses well-defined statistical equilibria in all cases.

Let us consider $\Feq$ a local extremum of entropy. Lagrange multipliers then guarantee the existence of a function ${ \alpha (\bK) }$, a scalar $\beta$ and a vector $\bgamma$ that ensure the differential equality
\begin{equation}
	\mD_{\Feq\!} S \!+\! \!\!\int\!\! \rd \bK \, \alpha (\bK) \mD_{\Feq\!} N(\bK) \!-\! \beta \mD_{\Feq\!} \Etot \!+\! \bgamma \!\cdot\! \mD_{\Feq\!} \bLtot \!=\! 0 .
	\label{eq:Lagrange_multipliers}
\end{equation}
In that expression, the differentials are given by the linear forms
\begin{align}
	{} & \mD_{\Feq\!} S \!: \delta F \mapsto - \kB \!\!\int\!\! \rd \hbL \rd \bK \, \big[ 1 \!+\! \ln \!\big( \Feq (\hbL , \bK) \big) \big] \, \delta F(\hbL, \bK) ,
	\nonumber
	\\
	{} & \mD_{\Feq\!} N(\bK) \!: \delta F \mapsto \!\!\int\!\! \rd \hbL \, \delta F(\hbL, \bK) ,
	\nonumber
	\\
	{} & \mD_{\Feq\!} \Etot \!: \delta F \mapsto \!\! \int \!\! \rd \hbL \rd \bK \, \veps (\hbL , \bK) \, \delta F (\hbL , \bK) ,
	\nonumber
	\\
	{} & \mD_{\Feq\!} \bLtot \!: \delta F \mapsto \!\!\int\!\! \rd \hbL \rd \bK \, L(\bK) \, \hbL \, \delta F(\hbL, \bK) ,
	\label{eq:explicit_Differentials}
\end{align}
where we used the symmetry ${ \mH_{\ell} [\bK , \bKp] \!=\! \mH_{\ell} [\bKp , \bKp] }$ (see equation~\ref{def:J_l}) to compute ${ \mD_{\Feq\!} \Etot }$. Injecting these relations into equation~\eqref{eq:Lagrange_multipliers}, we find
\begin{align}
	\forall \, \delta F, \; \!\!\int\!\! \rd \hbL \rd \bK \, \big\{
	 \! {} & - \kB \big[ 1 + \ln \! \big( \Feq (\hbL , \bK) \big) \big] + \alpha (\bK) 
	\label{eq:full_differential}
	\\
	{} & - \beta \, \veps(\hbL , \bK) + L(\bK) \, \bgamma \!\cdot\! \hbL \big\} \, \delta F (\hbL , \bK) = 0 ,
\nonumber
\end{align}
Since this integral must vanish whatever the small displacement ${ \delta F }$ considered, we have
\begin{align}
	\forall \, \hbL, \forall \, \bK, {} & - \kB \big[ 1 + \ln \! \big( \Feq (\hbL , \bK) \big) \big] + \alpha (\bK)
	\nonumber
	\\
	{} & + \beta \, \veps(\hbL , \bK) + L(\bK) \, \bgamma \!\cdot\! \hbL = 0 .
	\label{eq:Condition_for_extremum}
\end{align}
Inverting this relation, we find that the equilibrium \DF\ is necessarily of the form
\begin{equation}
	\Feq(\hbL, \bK) = \exp \!\big[ 
	\big( \alpha (\bK) \!-\! 1 \big) 
	\!-\! \beta \, \veps(\hbL , \bK) 
	\!+\! L(\bK) \, \bgamma \!\cdot\! \hbL \big] ,
	\label{eq:Shape_From_Lagrange}
\end{equation}
where, for convenience, the factor $\kB$ has been absorbed in the definition of $\alpha$, $\beta$ and $\bgamma$. Following equation~\eqref{def_NK}, we impose the number density of stars with orbital parameters $\bK$ to be ${ N (\bK) }$, and we finally recover equation~\eqref{eq:Shape_from_Lagrange}.

\section{Consistency functions and Jacobian}
\label{sec:ConsistencyJac}

We detail here the expressions of the consistency functions from equation~\eqref{def_consistency}. In order to shorten the notations, let us first write the exponential factor from equation~\eqref{eq:Shape_from_Lagrange} as
\begin{equation}
	\re [\hbL , \bK] = \exp \!\big[ \!-\! \beta \, \veps (\hbL , \bK) + \gamma \, L (\bK) \, \hLz \big] .
	\label{short_exp}
\end{equation}
We may then introduce the four angular integrals
\begin{align}
	I (\bK) 
	{} & = \!\! \int \!\! \rd \hbL \, \re [\hbL , \bK] ,\nonumber \\
	I_{\veps} (\bK) 
	{} & = \!\! \int \!\! \rd \hbL \, \veps (\hbL , \bK) \, \re [\hbL , \bK] , \nonumber \\
	I_{L} (\bK) 
	{} & = \!\! \int \!\! \rd \hbL \, L (\bK) \, \hLz \, \re [\hbL , \bK] , \nonumber \\
	I_{\ell} (\bK) 
	{} & = \!\! \int \!\! \rd \hbL \, Y_{\ell 0} (\hbL) \, \re [\hbL , \bK] .
	\label{def_Int_consistency}
\end{align}
The consistency functions from equation~\eqref{def_consistency} then simply read
\begin{align}
	C_{E} 
	{} & = \frac{1}{\Etot} \bigg\{ \Etot - \half \!\! \int \!\! \rd \bK \, N (\bK) \, \frac{I_{\veps}(\bK)}{I (\bK)} \bigg\} , \nonumber \\
	C_{L} 
	{} & = \frac{1}{\Ltot} \bigg\{ \Ltot - \!\! \int \!\! \rd \bK \, N (\bK) \, \frac{I_{L} (\bK)}{I (\bK)} \bigg\} , \nonumber \\
	C_{\Mellk} {} & = \frac{1}{N_{k} y_{\ell}} \bigg\{ \Mellk - N_{k} \frac{I_{\ell} (\bK)}{I (\bK)} \bigg\} .
	\label{consistency_explicit_app}
\end{align}
As required by equation~\eqref{iteration_Newton}, it is straightforward to compute the gradients of ${ (C_{E} , C_{L} , \{ C_{\Mellk} \}) }$ w.r.t.\ the order parameters ${ (\beta , \gamma , \{ \Mellk \}) }$. All these expressions are analytical and involve angular integrals similar to the ones present in equation~\eqref{def_Int_consistency}. In practice, having discretised the stellar populations (see~\S\ref{sec:DiscK}), the integrals w.r.t.\ ${ \rd \bK }$ simply become ${ \sum_{k} }$. In addition, benefiting from the assumption of axisymmetry (see~\S\ref{sub:AxiAssump}), the integrals w.r.t.\ ${ \rd \hbL }$ become integrals w.r.t.\ ${ \rd \hLz }$, which are computed using a Gauss--Legendre quadrature~\citep[see, e.g.,\@][]{Press+2007} with ${ 100 }$ nodes.

\section{Optimisation strategies}
\label{sec:caloric}

In this Appendix, we present our two main approaches to solving the iteration problem from equation~\eqref{iteration_Newton}.

\subsection{Random initialisations}
\label{sec:Random_Init}

A first approach to initialise equation~\eqref{iteration_Newton} is to consider starting points $\bT_{0}$ taken at random. More precisely, we draw the initial order parameters independently from one another and uniformly within the domains
\begin{align}
	{} & 0 \leq \beta \leq 20 \, \beta_{0}
	\quad \text{with} \quad
	\beta_{0} = 1 / \big( N \, G \mmin^{2} / \amin \big) , \nonumber \\ 
	{} & 0 \leq \gamma \leq 2 \, \gamma_{0}
	\quad \;\; \text{with} \quad
	\gamma_{0} = 1 / \big( \mmin \sqrt{G \MBH \amin} \big) , \nonumber \\
	{} & | \Mellk | \leq y_{\ell} \, N_{k} ,
	\label{def_range_random}
\end{align}
with ${ y_{\ell} \!=\! \sqrt{(2 \ell \!+\! 1)/(4 \pi)} }$.
The main advantage of such an agnostic approach is that it allows us to recover a cluster's both stable and unstable equilibria.

\begin{figure}
	\centering
	\includegraphics[width=0.45 \textwidth]{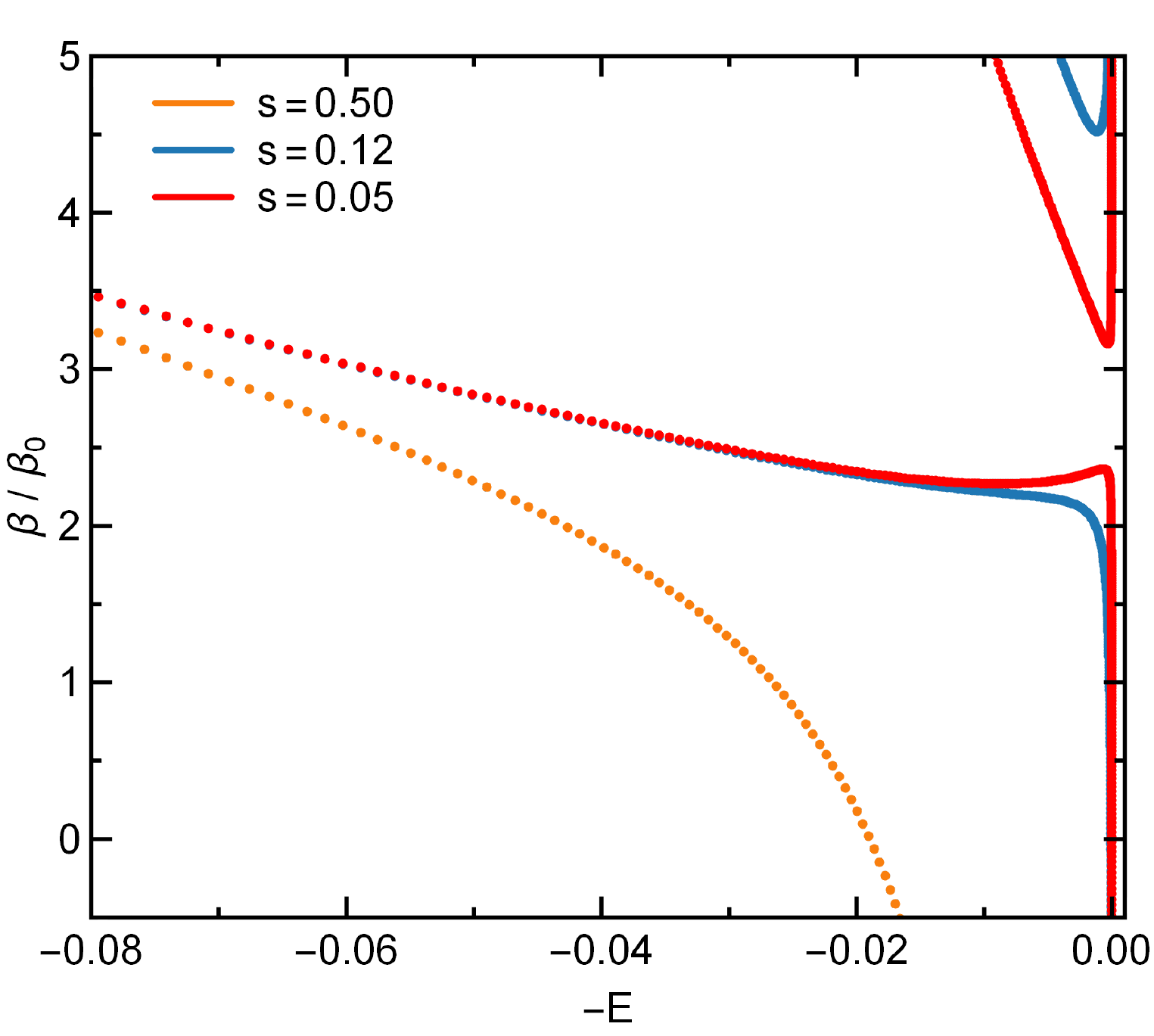}
	\caption{Inverse temperature $\beta$ as a function of the normalised total energy, ${ -E }$ (see equation~\ref{def_E_s}), for various spins $s$, when considering a cluster composed of $5$ populations of different individual masses interacting with ${ \ellmax \!=\! 10 }$. This figure was obtained without restricting ourselves to the global thermodynamical equilibria (see~\ref{sec:Random_Init}), as emphasised by the second set of (red and blue) solutions at high $\beta$.}
	\label{fig:Caloric_curves}
\end{figure}

In Fig.~\ref{fig:Caloric_curves}, we use this protocol for a multi-population and a multi-harmonic cluster with ${ (N_{m} , N_{a} , N_{e}) \!=\! (5,1,1) }$, and ${ \ellmax \!=\! 10 }$ -- see~\S\ref{sec:NumericalDetails} for details on the cluster's properties.
In a sense, this figure is similar to the top panel of Fig.~\ref{fig:Series_of_energies}, except that in Fig.~\ref{fig:Caloric_curves}, we do not restrict ourselves solely to the global thermodynamical equilibria. Interestingly, it also appears that all the qualitative behaviours reported in~\cite{Roupas+2017} and~\cite{Takacs+2018} are still present in the multi-population case, as can be seen by comparing with Figs.~\ref{fig:Roupas_fig_11} and~\ref{fig:Takacs_fig_4}.

We first note that for any spin $s$, there exists a region of the caloric curve where equilibria have negative temperatures. Negative temperatures occur when the total volume of phase space is finite (which is the case for \VRR\@), as first shown by~\cite{Onsager1949} in the context of ${2D}$ vortex dynamics. However, in the present context, the negative temperature states do not seem to have any particular impact and, furthermore, the region ${ \beta \!<\! 0 }$ is found to be outside of the astrophysically relevant regime (see Fig.~\ref{fig:Histogram}).

For some binding energies and spins, there can exist several solutions to equation~\eqref{eq:Shape_from_Lagrange}. Indeed, the Lagrange multipliers method finds all local extrema of the entropy. These could be local minima, saddle points, metastable local maxima, or the one global maximum. In the present context, the turning point method of Poincar\'e~\citep[see, e.g.\@,][]{LyndenBell+1968,Katz1978} states that along a continuous branch in the ${ (\Etot , \beta) }$-plane, there can be a loss of stability around a given solution in the microcanonical ensemble (i.e.\ fixed $\Etot$ and $\Ltot$) only if ${ \Etot \!\mapsto\! \beta (\Etot) }$ exhibits an infinite derivative there. In Fig.~\ref{fig:Caloric_curves}, no such turning points are observed. Given that the solution for ${ \beta \!=\! 0 }$ is always found to be stable (see~\S\ref{sec:beta_zero}), we can conclude that the branch of solutions that goes through ${ \beta \!=\! 0 }$ are always, at least, metastable. In practice, we systematically computed the entropy of the various branches, and the one going through ${ \beta \!=\! 0 }$ was always found to correspond to the global (axisymmetric) maxima. This justifies our choice in~\S\ref{sec:Iterative_resolutions} to limit ourselves to only determining the equilibria along ${ \beta \!=\! 0 }$, therefore significantly alleviating the numerical burden.

In Fig.~\ref{fig:Caloric_curves}, we also recover that low-spin multi-population clusters can exhibit negative specific heats at small energies. However, when considering invariants ${ (E,s) }$ provided by the protocol from~\S\ref{sub:param}, none of the realisations considered were found to fall within that domain. Finally, we note that the region of negative specific heat is delimited by a pair of equilibrium points where the function ${ \Etot \!\mapsto\! \beta (\Etot) }$ has a vanishing derivative. The Poincar\'e turning point method can be used to show that the series of equilibria in between these two points of vanishing derivative is unstable in the canonical ensemble (fixed $\beta$ and $\gamma$). This leads to a situation of ensemble inequivalence -- since equilibria with negative specific heats are allowed in the microcanonical ensemble but forbidden in the canonical ensemble -- characteristic of systems driven by long-range interactions~\citep[see, e.g.\@,][]{Chavanis+2006}. Given the similarities of Fig.~\ref{fig:Caloric_curves} with the results already presented in~\cite{Roupas+2017,Takacs+2018}, we refer to these previous works for detailed discussions of these canonical phase transitions. We stress, however, that the relevant ensemble considered in the main text is the microcanonical one since our system is assumed to be isolated.

\subsection{Iterative resolutions}
\label{sec:Iterative_resolutions}

When dealing with systems with a much larger number of stellar populations, the random initialisation from equation~\eqref{def_range_random} is not efficient anymore. Indeed, because the total number of order parameters scale like ${ \mO (\Npop \, \ellmax) }$ (see equation~\ref{def_bT}), the likelihood of starting sufficiently close to one equilibrium for equation~\eqref{iteration_Newton} to converge drastically drops. As such, benefiting from the insight of Fig.~\ref{fig:Caloric_curves}, we limit ourselves to predicting the cluster's global thermodynamical equilibrium, i.e.\ predicting the equilibrium that lies on the branch that goes through ${ \beta \!=\! 0 }$.

More precisely, assuming that the cluster at hand is characterised by the two invariants ${ (\Etot , \Ltot) }$, we first determine the energy $\Etot^{(0)}$ and order parameters $\bT^{(0)}$ associated with the thermodynamical equilibrium at temperature ${ \beta \!=\! 0 }$ and total angular momentum $\Ltot$. We highlight in~\S\ref{sec:beta_zero} how such a problem is significantly easier to solve, as it does not involve any self-gravitating contribution, therefore making equation~\eqref{eq:Shape_from_Lagrange} explicit.

From this initial configuration, we use $\bT^{(0)}$ as an initial condition to solve a new self-consistency problem for the invariants ${ ( \Etot^{(1)} , \Ltot ) }$, with $\Etot^{(1)}$ lying between $\Etot^{(0)}$ and the final target $\Etot$. Provided that $\Etot^{(1)}$ is close enough to $\Etot^{(0)}$ the Newton method from~\S\ref{sub:SelfCons} is expected to converge rapidly and to provide us with a new configuration $\bT^{(1)}$, solution of the problem with invariants ${ (\Etot^{(1)} , \Ltot) }$. This new solution may then be used as an appropriate initial condition to solve the problem ${ (\Etot^{(2)} , \Ltot) }$, getting us closer to our target energy $\Etot$. Repeating this iterative process, we can ultimately solve the problem truly at hand, i.e.\ the one associated with the invariants ${ (\Etot , \Ltot) }$. In practice, we generically used a total of ${50}$ different energies ${ \Etot^{(i)} }$ spread logarithmically between $\Etot^{(0)}$ and the target total energy $\Etot$.

\subsection{Equilibria for \texorpdfstring{${ \beta = 0 }$}{beta = 0}}
\label{sec:beta_zero}

We derive the \DF\ that solves the entropy maximisation problem, without any constraint on energy. This \DF\ is the starting point of our optimisation strategy with small steps in total energy (see~\S\ref{sec:Iterative_resolutions}).

First, for a given non-zero value of $\bLtot$ (taken to be along ${+z}$), the solution of the optimisation problem without energy constraint must be axisymmetric. Indeed, if ${ F (\phi , \hLz , \bK) }$ satisfies the constraints from equations~\eqref{def_NK} and~\eqref{def_Ltot}, then so does ${ F_{\Delta} : (\phi, \hLz, \bK) \!\rightarrow\! F (\phi \!+\! \Delta, \hLz , \bK) }$. Given that equation~\eqref{def_Ltot} is linear w.r.t. $F$, 
\begin{equation} 
	\oF : (\phi, \hLz , \bK) \rightarrow \!\!\int_{0}^{2\pi} \!\! \frac{\rd \Delta}{2 \pi} \, F_{\Delta} (\phi, \hLz , \bK) = \oF (\hLz, \bK)
\label{def:oF}
\end{equation}
also meets the constraints on $\bLtot$. We note that $\oF$ is a barycentre of the phase-shifted \DFs\@, which all have the same entropy (equation~\ref{def_S}). Given that ${ s \!:x \!\mapsto\! x \ln (x) }$ is a convex function, Jensen's inequality gives ${ S[\oF] \!\geq\! S(F) }$. As such, in the absence of any constraint on energy, the entropy maximum must be axisymmetric.

Given this axisymmetry and remembering that ${ \beta \!=\! 0 }$ since there is no constraint on the energy, equation~\eqref{eq:Shape_from_Lagrange} becomes
\begin{equation}
	\Feq(\hbL, \bK) = N(\bK) \, \frac{ \exp \!\big[ L(\bK) \, \gamma \, \hLz \big]}{
	\displaystyle \int \!\! \rd \hbLp \exp \!\big[ L(\bK) \, \gamma \, \hLp_{z} \big]} .
	\label{eq:Shape_from_Lagrange_no_energy_constraint}
\end{equation}
Equation~\eqref{eq:Shape_from_Lagrange_no_energy_constraint} is much simpler than equation~\eqref{eq:Shape_from_Lagrange} because its r.h.s.\ does not contain $\Feq$ anymore, i.e.\ the equation becomes explicit. 

To find a value of $\gamma$ for which $\Feq$ meets the constraint on $\Ltot$, we compute the integral on the r.h.s.\ of equation~\eqref{def_Ltot}. We get
\begin{align}
	L(\bK) \!\! \int \!\! \rd \hbL \, {} & \hLz \exp \!\big[ L(\bK) \, \gamma \, \hL_{z} \big]
	= 2 \pi \, L(\bK) \!\!\int_{-1}^{+1} \!\!\!\!\!\!\! \rd z \, z \, \re^{L(\bK) \gamma \, z} \nonumber \\
	= {} & \frac{4 \pi}{\gamma} \big\{ \cosh [ L(\bK) \, \gamma ] - \sinch [L(\bK) \, \gamma ] \big\} ,
	\label{eq:Internal_integral_no_energy_constraint}
\end{align}
and
\begin{align}
	\int \!\! \rd \hbL \exp \!\big[ L(\bK) \, \gamma \, \hL_{z} \big] 
	= {} & 2 \pi \!\!\int_{-1}^{+1} \!\!\!\!\!\!\! \rd z \, \re^{ L(\bK) \gamma \, z } \nonumber \\
	= {} & 4 \pi \, \sinch [ L(\bK) \, \gamma ] ,
	\label{eq:Normalisation_no_energy_constraint}
\end{align}
with ${ \sinch(x) \!=\! \sinh(x)/x }$.
Therefore, equation~\eqref{def_Ltot} finally gives
\begin{equation}
	\Ltot(\gamma) = \!\! \int \!\! \rd \bK \, N(\bK) \big\{ L(\bK) \, \cotanh [ L(\bK) \, \gamma] - (1/\gamma) \big\} .
	\label{eq:ConstraintL_no_energy_case}
\end{equation}
This is a strictly increasing function of $\gamma$, so that there exists at most one solution $\gamma$ for a given value of $\Ltot$. Although not analytical, such a solution is straightforward to obtain by dichotomy.

\section{Numerical applications}
\label{sec:NumericalDetails}

We briefly detail some of our choices in the effective numerical implementation of the method of entropy optimisation from~\S\ref{sub:SelfCons} We also validate it in single-population systems by reproducing previous published results.

\subsection{Discretisation of the stellar populations}
\label{sec:DiscK}

As explained in~\S\ref{sub:SelfCons}, in order to be effectively implemented the entropy optimisation requires a discretisation of the stellar populations. Each population is associated with an index $k$ and is characterised by some orbital parameters $\bK_{k}$ and a particle number $N_{k}$.

In~\S\ref{sub:param}, we assumed that ${ (m,a,e) }$ are drawn independently from one another, so that they may be discretised independently as well. For the stellar mass, the considered range ${ \mmin \!\leq\! m \!\leq\! \mmax }$ is discretised in ${ N_{m} }$ logarithmic bins. Similarly, for $a$ we use ${ N_{a} }$ logarithmic bins, and ${ N_{e} }$ linear bins for $e$. As a consequence, the effective number of populations is set by ${ \Npop \!=\! N_{m} N_{a} N_{e} }$. Then, for a given population, the number of particles $N_{k}$ is simply set by the value of the sampling power-law \PDFs\ in the centre of the bins, multiplied by the volumes of the bins. Similarly, the value of the population's orbital parameter, $\bK_{k}$, is set by the value of $\bK$ in the centre of the bin. As such, all our integrals over ${ \rd \bK }$ are formally approximated by Riemann sums using the midpoint rule.

To obtain Figs.~\ref{fig:SegregationMass} and~\ref{fig:Distribution_on_the_sphere}, we used ${ (N_{m} , N_{a} , N_{e}) \!=\! (20, 10 , 5) }$. For the subsequent sections, we considered the simpler regime ${ (N_{m} , N_{a} , N_{e}) \!=\! (10 , 10 , 3) }$. We checked that such choices of population numbers did not affect the results. Given that the numerical complexity scales like ${ \mO (\Npop^{3}) }$, this greatly eases the overall parameter exploration.

\subsection{Convergence w.r.t. \texorpdfstring{$\ellmax$}{ellmax}}
\label{sec:Convlmax}

In practice, the Hamiltonian from equation~\eqref{eq:Htot} has to be truncated to some finite harmonic order $\ellmax$. While~\cite{Roupas+2017} limited themselves to the quadrupolar case ${ \ellmax \!=\! 2 }$, \cite{Takacs+2018} lifted this restriction in single-population systems and showed that an effective truncation at ${ \ellmax \!=\! 10 }$ is sufficient to get converged results. This is what we briefly explore for multi-population clusters in this section.

\begin{figure}
	\centering
	\includegraphics[width= 0.45 \textwidth]{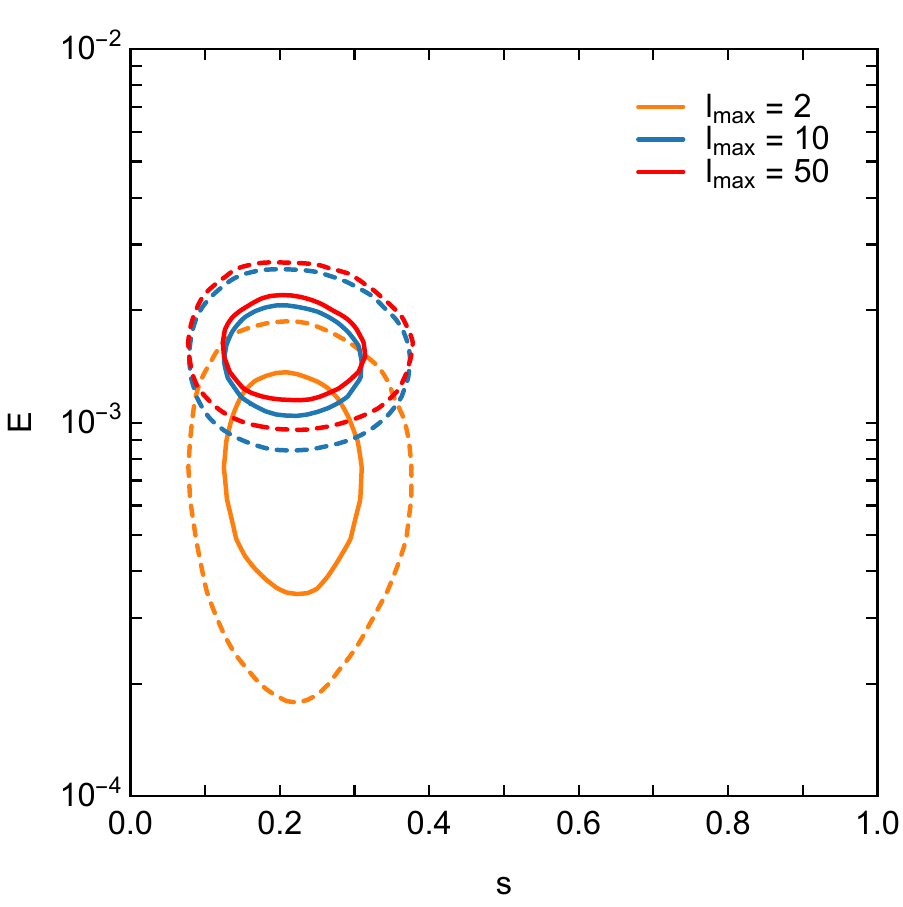}
	\caption{Illustration of the dependence of ${ (E,s) }$ as a function of $\ellmax$ for the protocol from~\S\ref{sub:param}. For a given value of $\ellmax$, the two plotted contours correspond to ${ 33\% }$ (dashed) and ${ 66\% }$ (solid) of the \PDF\@'s maximum. As already pointed out in~\protect\cite{Takacs+2018}, for ${ \ellmax \!\gtrsim\! 10 }$, the values of the two invariants ${ (E,s) }$ can be considered as converged. }
	\label{fig:Histo_lmax}
\end{figure}

First, in Fig.~\ref{fig:Histo_lmax}, considering the same initialisation protocol as in~\S\ref{sub:param}, we illustrate the dependence of the invariants ${ (E,s) }$ as a function of $\ellmax$.
In that figure, we recover that the typical value of $E$ increases with $\ellmax$, while, of course, the average value of $s$ is independent of it. In addition, we note that restricting oneself to ${ \ellmax \!=\! 2 }$ seems insufficient, while ${ \ellmax \!=\! 10 }$ offers reasonably well converged values of the binding energy.

\begin{figure}
	\centering
	\includegraphics[width= 0.45 \textwidth]{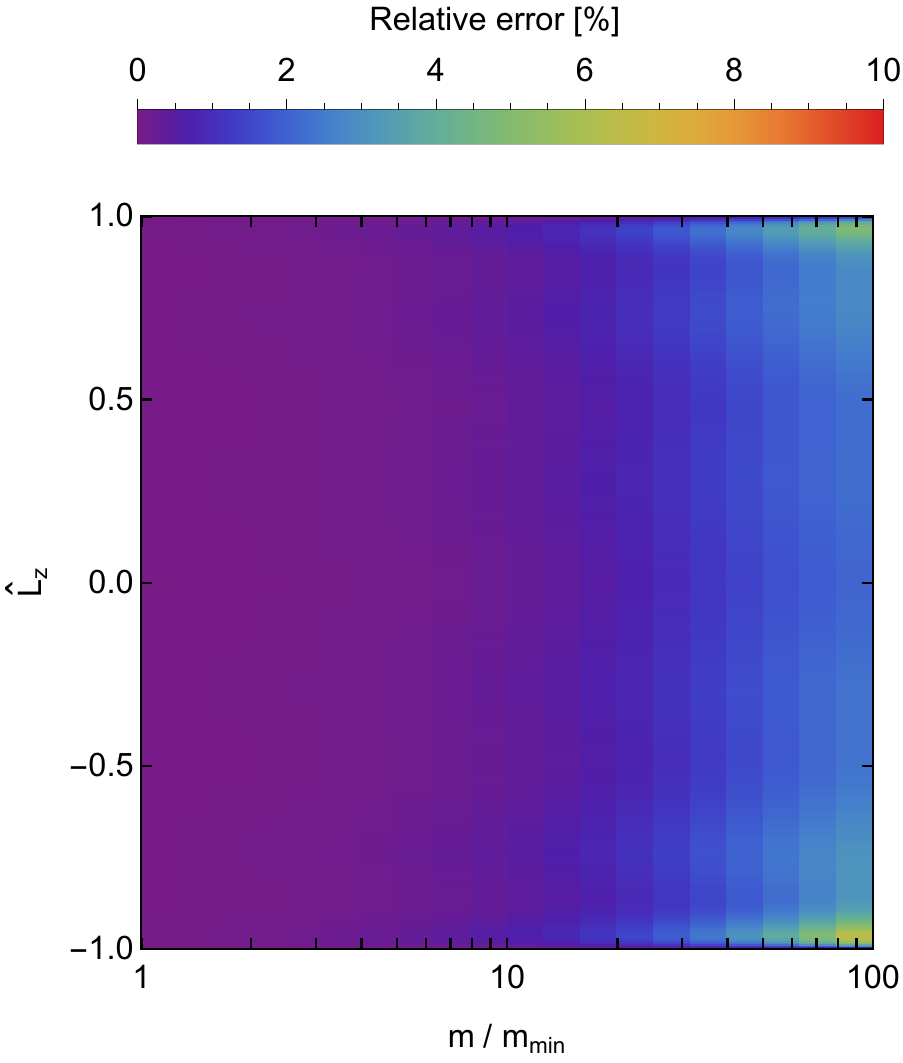}
	\caption{Relative error in Fig.~\ref{fig:SegregationMass} as one increases the harmonic truncation from ${ \ellmax \!=\! 10 }$ to ${ \ellmax \!=\! 50 }$. The maximum relative error is ${ \sim\! 7\% }$. }
	\label{fig:Szolgyen_lmax}
\end{figure}

In addition to affecting the values of the cluster's invariants, increasing the value of $\ellmax$ might also impact the overall shape of the reconstructed anisotropic equilibria \DF\@. This is what we explore in Fig.~\ref{fig:Szolgyen_lmax}, where we compute the relative error in the reconstructed \DF\ between ${ \ellmax \!=\! 10 }$ and ${ \ellmax \!=\! 50 }$.
Given that the maximum relative error is ${ \sim\! 7\% }$ for the present clusters, we systematically truncated the pairwise interaction at ${ \ellmax \!=\! 10 }$ in all the figures presented in the main text. As the complexity of the entropy optimisation scales like ${ \mO (\ellmax^{3}) }$, this significantly alleviates the numerical difficulty of the computations.

\subsection{Validation in single-population systems}
\label{sec:validation}

We validate our implementation of entropy maximisation by recovering previous results from the literature.

\begin{figure}
	\centering
	\includegraphics[width=0.45 \textwidth]{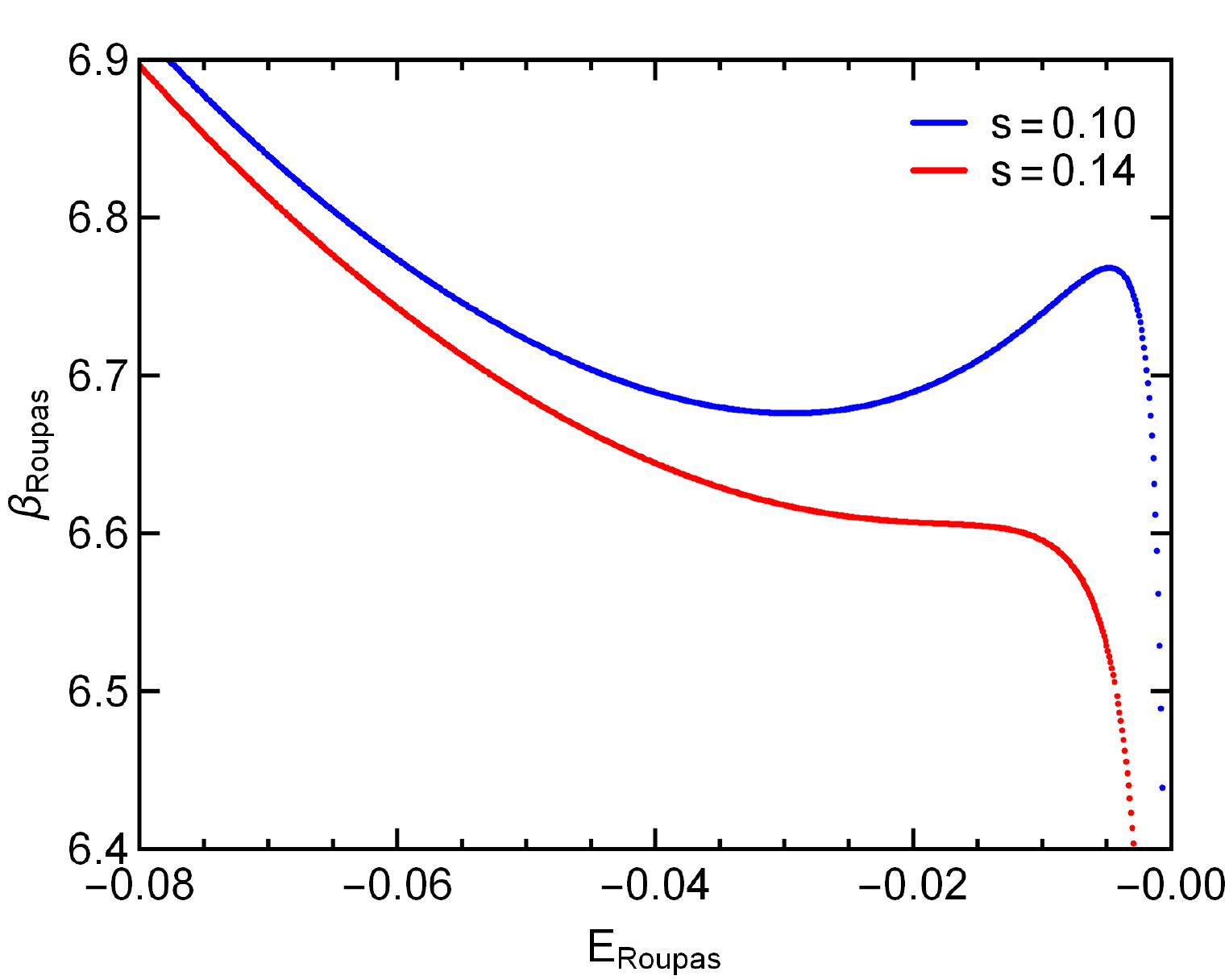}
	\caption{Inverse temperature $\beta$ as a function of the energy $E_{\mathrm{Roupas}}$, for various fixed total angular momentum, $s$, for a single-population cluster with ${ \ellmax \!=\! 2 }$. This figure reproduces fig.~{11} of~\protect\cite{Roupas+2017}. For low $s$, the cluster can exhibit a negative specific heat, i.e.\ ${ \p \beta_{\mathrm{Roupas}} / \p E_{\mathrm{Roupas}} \!>\! 0 }$. }
	\label{fig:Roupas_fig_11}
\end{figure}

First, in Fig.~\ref{fig:Roupas_fig_11}, we recover the caloric curve presented in fig.~{11} of~\cite{Roupas+2017} for a single-population system interacting only through ${ \ell \!=\! 2 }$.
In order to exactly match our present normalisation convention with the ones from~\cite{Roupas+2017}, one has to consider
\begin{equation}
	\beta_{\mathrm{Roupas}} \!=\! \beta \, \tfrac{3}{8} N G m^{2} \! / a ;
	\;
	E_{\mathrm{Roupas}} \!=\! \Etot / \big( \tfrac{3}{8} N^{2} G m^{2} \! / a \big) .
	\label{convention_Roupas}
\end{equation}
In this figure, it is interesting to recall that clusters with sufficiently small total angular momentum can present a negative specific heat.

\begin{figure}
	\centering
	\includegraphics[width= 0.45 \textwidth]{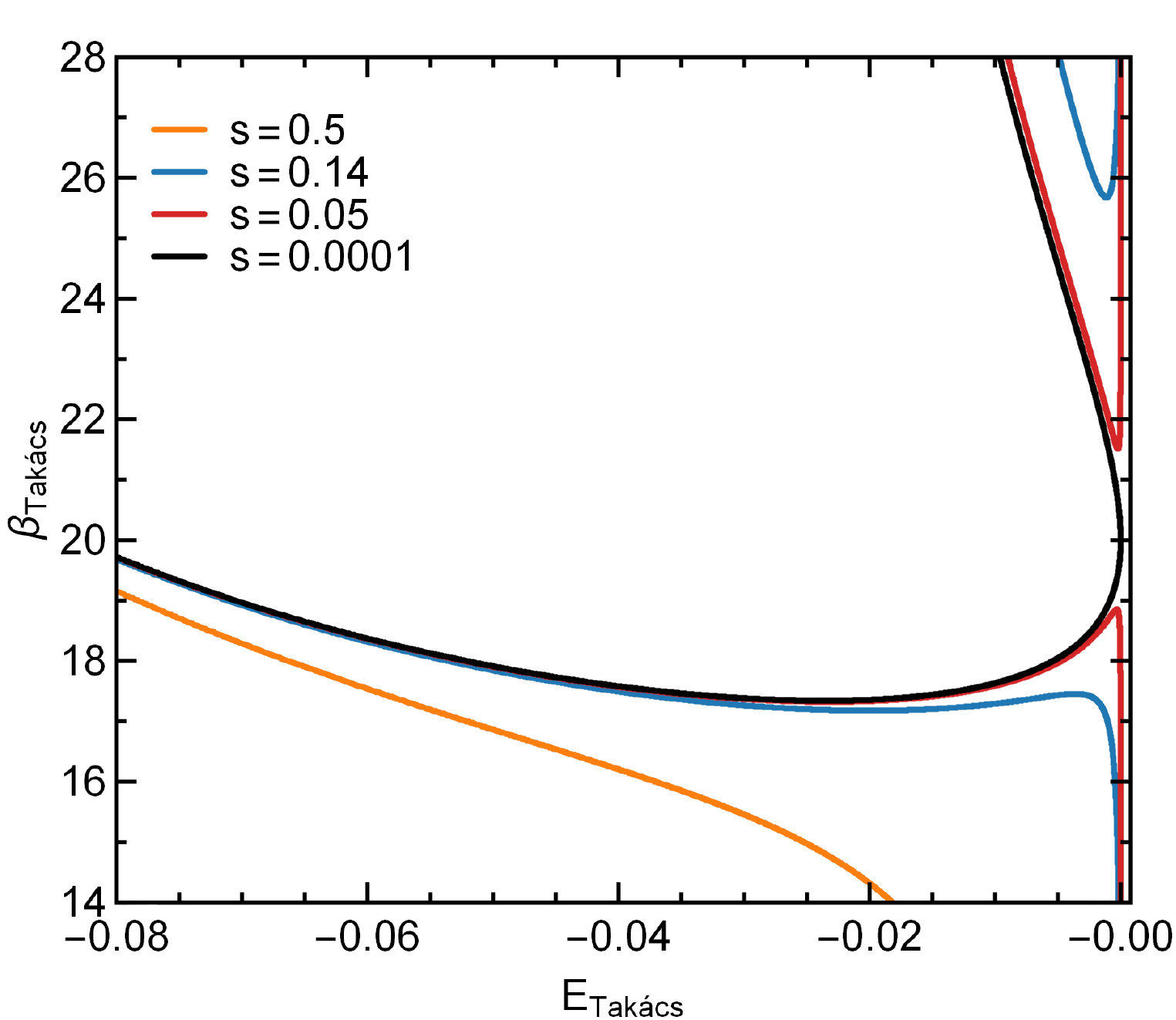}
	\caption{Same as Fig.~\ref{fig:Roupas_fig_11} but considering ${ \ellmax \!=\! 58 }$. This figure reproduces fig.~{4} of~\protect\cite{Takacs+2018}. In addition to the presence of negative specific heats for low $s$, we recover for weakly bound clusters the presence of more than one equilibrium solutions. Configurations with the smallest $\beta$ were always found to have the highest entropy, hence correspond to the cluster's (axisymmetric) global statistical equilibria.}
	\label{fig:Takacs_fig_4}
\end{figure}

In Fig.~\ref{fig:Takacs_fig_4}, we recover fig.~{4} from~\cite{Takacs+2018}, for a single-population system coupled beyond the quadrupolar interaction.
To exactly match the normalisation conventions from~\cite{Takacs+2018}, one has to consider
\begin{equation}
	\beta_{\mathrm{Tak\acute{a}cs}} = \beta \, N G m^{2} \!/ a ;
	\quad
	E_{\mathrm{Tak\acute{a}cs}} = \Etot / \big( N^{2} G m^{2} \!/ a \big) .
	\label{convention_Takacs}
\end{equation}
In addition, following equation~{(2)} of~\cite{Takacs+2018}, we also had to replace the coupling coefficients from equation~\eqref{def:J_l} with the simpler asymptotic scaling
\begin{equation}
	\mH_{\ell}^{\mathrm{Tak\acute{a}cs}} = \frac{G m^{2}}{a} \frac{4 \pi}{\ell^{2} \, (2 \ell + 1)} .
	\label{scaling}
\end{equation}
For sufficiently small values of the total spin $s$, one recovers more than one equilibrium solutions. The solutions with the smallest $\beta$, i.e.\ the branch that goes through ${ \beta \!=\! 0 }$, were always found to have the largest entropy, hence corresponding to the (axisymmetric) thermodynamical equilibria.

\end{document}